\documentclass[12pt]{iopart}
\usepackage{graphicx}
\usepackage{setspace}
\usepackage{texdraw}
\font\mybbb=msbm10
\def\Bbb#1{\mbox{\mybbb #1}}
\begin{document}
\title{Entropy of  $XY$ Spin Chain and  \\
Block Toeplitz Determinants}

\author{A. R. Its\dag, \  \ B.-Q. Jin\S  \  \ and V. E. Korepin $\star $ }

\address{\dag\ Department of Mathematical Sciences, Indiana
University-Purdue University Indianapolis, Indianapolis, IN
46202-3216, USA}
\address{\S\ College of Physics and Electronic Information, Wenzhou
University, Wenzhou, Zhejiang, P.R. China }
\address{$\star $\ C.N.\ Yang Institute for Theoretical Physics, State
University of New York at Stony Brook, Stony Brook, NY 11794-3840,
USA}

\ead{itsa@math.iupui.edu,jinbq@wzu.edu.cn,korepin@insti.physics.sunysb.edu}


\begin{abstract}
We consider  entanglement in the ground state of the  $XY$ model
on infinite chain. We use von Neumann  entropy of a sub-system as a measure of
entanglement.  The entropy of a large block of neighboring spins
approaches a constant as the size of the block increases. We
evaluate  this limiting  entropy as a function of anisotropy and
transverse magnetic field. We use
integrable Fredholm operators and the Riemann-Hilbert approach. The entropy reaches minimum at highly ordered states but increases boundlessly at phase transitions.
\end{abstract}

\maketitle

\section{Introduction}

Entanglement is a primary resource for quantum computation and
 information processing \cite{BD,L, ben,pop,dev,wint}.
It is necessary for quantum  control.
It shows how much quantum effects we can use to control one system by another.
Stable and large scale
entanglement is necessary for  scalability of  quantum
computation \cite{rasetti, zanardi, GRAC}. Entropy of a subsystem as a measure of entanglement was discovered in \cite{ben}.
Essential progress is achieved in understanding of  entanglement in
various quantum systems
\cite{fazio,nielsen,V,jin,LRV,K,cardy,ABV,VMC,rasetti,zanardi,honk,LO,PP,FS,salerno,fan, jvid,
kais,eisert1,eisert2,julien,plenio, briegel, bruno,fan}.

 $XY$ model in a transverse magnetic field was studied from the
point of view of quantum information in
\cite{vidal,fazio,keat,yang}.
In this paper we evaluate the entropy of a  block of $L$ neighboring
spins in the ground state of the $XY$ model in the limit
$L\rightarrow \infty $ analytically. Our approach uses representation of \cite{vidal} and  the
Riemann-Hilbert method of the theory of integrable Fredholm
operators. The final answer is given in terms of the elliptic
functions and is presented in Eq.~(\ref{3333May}) and (\ref{33May})
below.
The Hamiltonian of the $XY$  model can be written as
\begin{eqnarray}
{\cal H}=-\sum_{n=-\infty}^{\infty}
(1+\gamma)\sigma^x_{n}\sigma^x_{n+1}+(1-\gamma)\sigma^y_{n}\sigma^y_{n+1}
+ h\sigma^z_{n} \label{xxh}
\end{eqnarray}
Here $\gamma$ is the anisotropy parameter; $\sigma^x_n$,
$\sigma^y_n$ and  $\sigma^z_n$ are the Pauli matrices and $h$ is the
magnetic field. The  model was solved  in
\cite{Lieb,mccoy,mccoy2,gallavotti}. The methods of Toeplitz
determinants, as well as the techniques based on integrable Fredholm
operators, were used for the evaluation of some correlation
functions, see  \cite{mccoy2,aban,sla, dz, izer, pron}.

We consider  the ground state $|GS\rangle$ of the model. We evaluate
the entropy of a sub-system of the ground state. We shall calculate the
entropy of a block    of $L$ neighboring spins, it
measures the entanglement between the block  and the rest of the
chain \cite{ben}. We treat the whole chain as a binary system
$|GS\rangle = |A \& B\rangle $. We denote this block of $L$
neighboring spins by subsystem A and the rest of the chain by
subsystem B. The density matrix of the ground state can be denoted
by \mbox{$\rho_{AB}=|GS\rangle \langle GS|$}. The density matrix of
subsystem A is \mbox{$\rho_A= Tr_B(\rho_{AB})$}. The von Neumann
entropy $S(\rho_A)$  of subsystem A can be represented as follows:
\begin{eqnarray}
S(\rho_A)=-Tr_A(\rho_A \ln \rho_A) \qquad  \qquad \bowtie \label{edif} \label{olds}
\end{eqnarray}
This entropy also defines the dimension of the  Hilbert space of
states of subsystem A.

A set of Majorana operators were used in \cite{vidal} with
self-correlations described by the following matrix:
\begin{eqnarray}
\mathbf{B}_L=\left( \begin{array}{cccc}
\Pi_0 &\Pi_{-1}& \ldots &\Pi_{1-L}\\
\Pi_{1}& \Pi_0&   &   \vdots\\
\vdots &      & \ddots&\vdots\\
\Pi_{L-1}& \ldots& \ldots& \Pi_0
\end{array}     \right) \nonumber
\end{eqnarray}
Here
$$
\Pi_l=\frac{1}{2\pi} \int_{0}^{2\pi} \, \mathrm{d} \theta\,
e^{-\mathrm{i} l \theta} {\cal G}(\theta),\quad {\cal
G}(\theta)=\left( \begin{array}{cc}
               0& g(\theta)\\
               -g^{-1}(\theta)&0
               \end{array} \right)
               $$
\begin{equation}\textrm{and} \qquad g(\theta)=\frac{\cos \theta
-\mathrm{i} \gamma\sin \theta -h/2}{|\cos \theta -\mathrm{i}
\gamma\sin \theta
 -h/2|} \quad . \end{equation}

One can use an orthogonal matrix  $V$ to transform $\mathbf{B}_L$ to
a canonical form:
\begin{eqnarray}
V \mathbf{B}_{L} V^T= \oplus_{m=1}^{L} \nu_m \left(
\begin{array}{cc}
               0& 1\\
               -1&0
               \end{array} \right),\label{vd}
               \end{eqnarray}
               The real numbers  $-1<\nu_m<1$  play an important role.
               We shall call them eigenvalues.
               The entropy of a block of  $L$ neighboring
               spins was represented  in \cite{vidal} as
\begin{eqnarray}
S(\rho_A)&=&\sum_{m=1}^{L} H(\nu_m) \label{eaap1}
\end{eqnarray}
with
\begin{eqnarray}
 H( \nu)= -\frac{1+\nu}{2} \ln \frac{1+\nu}{2}-\frac{1-\nu}{2} \ln
\frac{1-\nu}{2}.\label{intee1}
\end{eqnarray}

\begin{figure}[ht]
\begin{center}
\includegraphics[width=3in,clip]{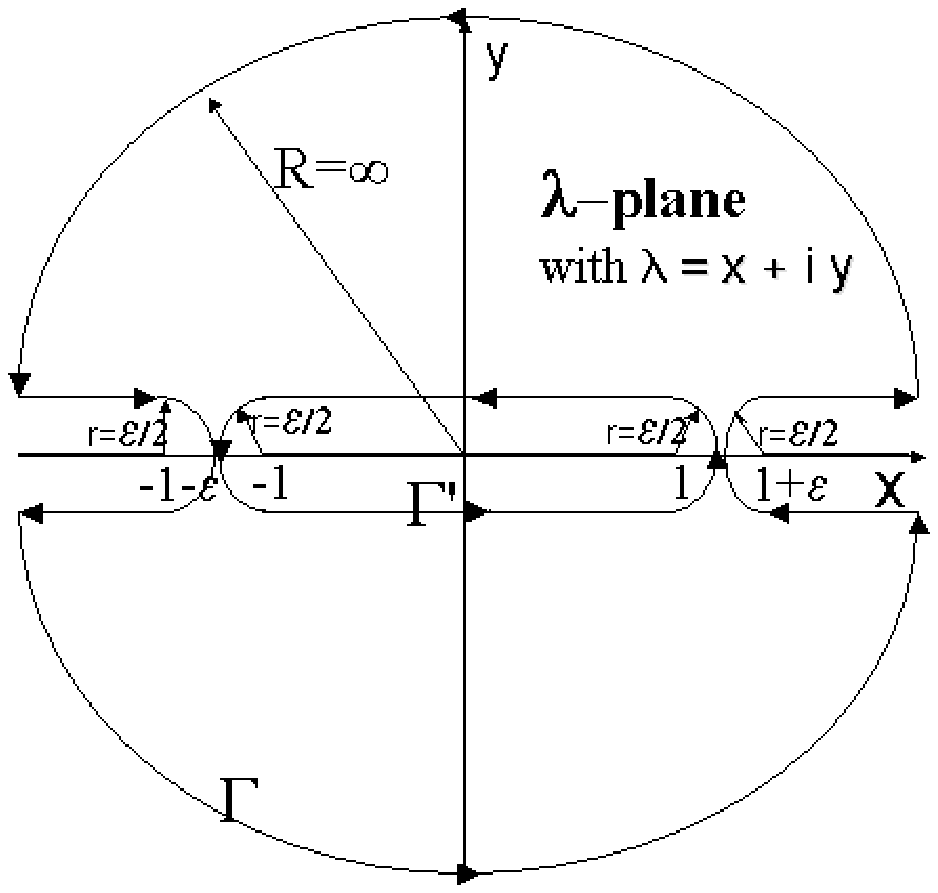}
\end{center}
\caption{\it Contours \mbox{$\Gamma'$} (smaller one) and
\mbox{$\Gamma $} (larger one). Bold lines $(-\infty, -1-\epsilon)$
and $(1+\epsilon,\infty)$ are the cuts of the integrand
$e(1+\epsilon,\lambda)$. Zeros of $D_{L}(\lambda)$ (Eq.~$\ref{exd}$)
are located on the bold line $(-1, 1)$. The arrows indicate the
directions of integrations, and $\mathrm{r}$ and $\mathrm{R}$ are
the radius of the circles. $\P $  } \label{fig1}
\end{figure}

In order to calculate the asymptotic form of the entropy it is not
convenient to use formula (\ref{vd}) and (\ref{eaap1}) directly.
Following the idea we have already used in \cite{jin}, let us
introduce:
\begin{eqnarray}
\widetilde{\mathbf{B}}_{L}(\lambda)=\mathrm{i}\lambda I_{L}-
\mathbf{B}_{L}, \quad D_{L}(\lambda)=\det
\widetilde{\mathbf{B}}_{L}(\lambda)
\end{eqnarray}
and \begin{eqnarray}
 e(x, \nu)= -\frac{x+\nu}{2} \ln \frac{x+\nu}{2}-\frac{x-\nu}{2} \ln
\frac{x-\nu}{2}.\label{intee}
\end{eqnarray}
 Here $I_{L}$ is the
identity matrix of the size $2L$. By definition, we have
$H(\nu)=e(1,\nu)$ and
\begin{eqnarray}
D_{L}(\lambda)=(-1)^{L} \prod_{m=1}^{L} (\lambda^2-\nu_m^2).
\label{exd}
\end{eqnarray}
With the help of the  Cauchy residue theorem  we rewrite formula
($\ref{eaap1}$)  in the following form:
\begin{eqnarray}
S(\rho_A)=\lim_{\epsilon \to 0^+} \frac{1}{4\pi \mathrm{i}}
\oint_{\Gamma'} \mathrm{d} \lambda\,  e(1+\epsilon, \lambda)
\frac{\mathrm{d}}{\mathrm{d} \lambda} \ln
D_{L}(\lambda)\; \qquad  \star  \label{eaa}
\end{eqnarray}
Here the contour \mbox{$\Gamma'$} is depicted in Fig~$\ref{fig1}$;
it encircles all zeros of \mbox{$D_{L}(\lambda)$}.
\noindent We also notice that $\widetilde{\mathbf{B}}_{L}(\lambda)$
is the block Toeplitz matrix,
\begin{eqnarray}
\widetilde{\mathbf{B}}_L(\lambda)=\left( \begin{array}{cccc}
\widetilde{\Pi}_0 &\widetilde{\Pi}_{-1}& \ldots
&\widetilde{\Pi}_{1-L}\\
\widetilde{\Pi}_{1}& \widetilde{\Pi}_0&   &   \vdots\\
\vdots &      & \ddots&\vdots\\
\widetilde{\Pi}_{L-1}& \ldots& \ldots& \widetilde{\Pi}_0
\end{array}     \right) \quad \textrm{with}\nonumber
\end{eqnarray}
\begin{equation}
 \widetilde{\Pi}_l=\frac{1}{2\pi\mathrm{i}}\oint_{\Xi} \,
\mathrm{d} z\, z^{-l-1} \Phi(z),
\end{equation}
where the matrix generator $\Phi(z)$ is defined by the equations,
\begin{equation}
 \Phi(z)=\left( \begin{array}{cc}
               \mathrm{i}\lambda & \phi(z)\\
               -\phi^{-1}(z)&\mathrm{i}\lambda
               \end{array} \right) \label{defphi}
\end{equation}
\begin{equation}
\textrm{and}\quad \phi(z)=
\left(\frac{\lambda_1^*}{\lambda_{1}}\frac{(1-\lambda_1\,
z)(1-\lambda_2\, z^{-1})}{(1-\lambda_1^* \, z^{-1})(1-\lambda_2^*\,
z)}\right)^{1/2}\label{defph}
\end{equation}

\begin{figure}[ht]
\begin{center}
\includegraphics[width=4in,clip]{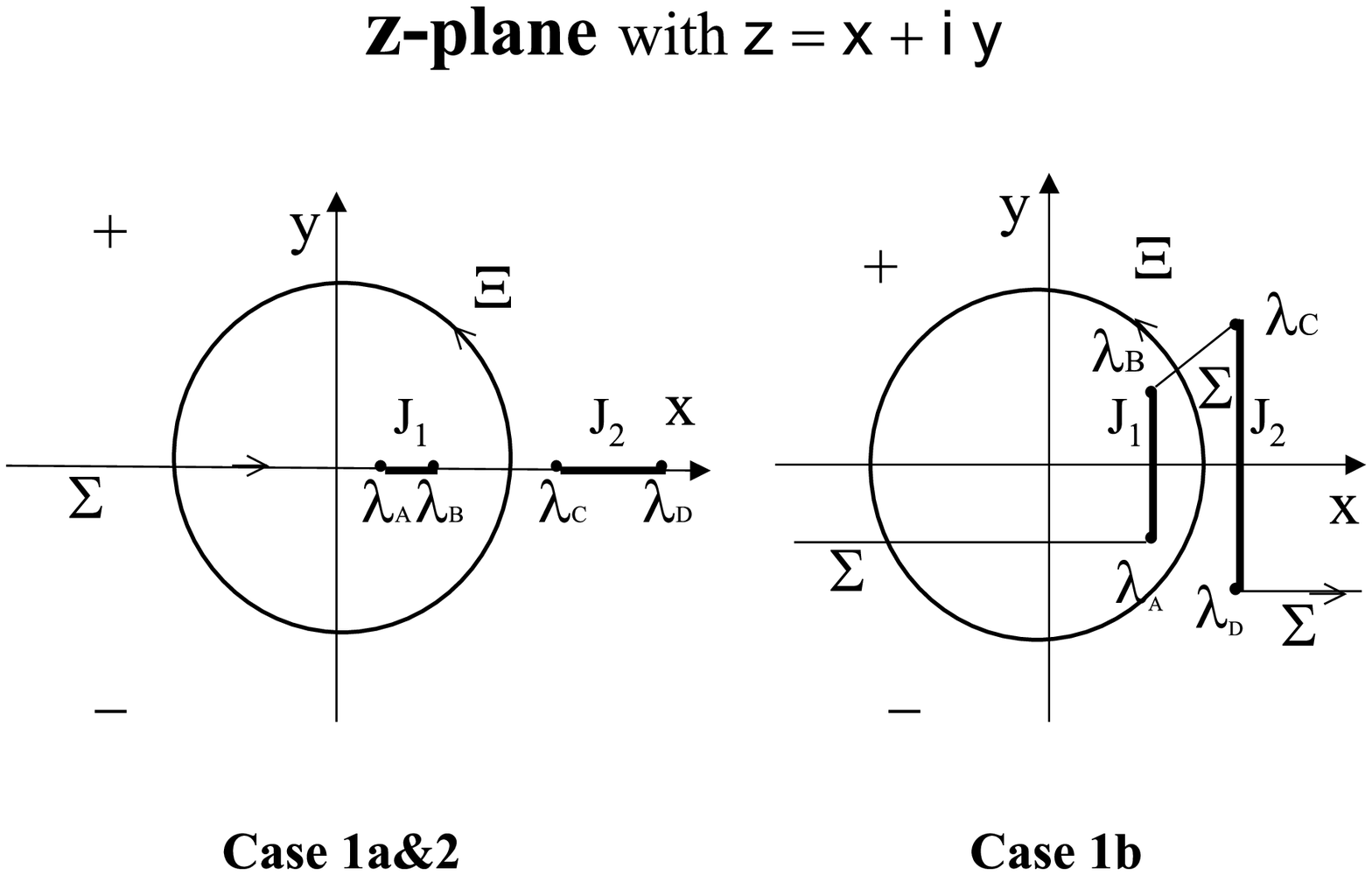}
\end{center}
\caption{\it The  polygonal line $\Sigma$ (oriented as it is
indicated) separates the complex $z$ plane into two parts: the part
$\Omega_{+}$ which lies to the left of $\Sigma$, and the part
$\Omega_{-}$ which lies to the right of $\Sigma$. Curve $\Xi$ is the
unit circle in anti-clockwise direction. Cuts $J_1, J_2$ for the
functions $\phi(z),w(z)$ are labeled by bold on the line $\Sigma$.
The definition of the end points of the cuts $\lambda_{\ldots}$
depends on the case: {\bf Case} $1$a: $\lambda_A=\lambda_1$,
$\lambda_B=\lambda_2^{-1}$, $\lambda_C= \lambda_2$, and $\lambda_D=
\lambda_1^{-1}$. {\bf Case} $1$b:
 $\lambda_A=\lambda_1$,
$\lambda_B=\lambda_2^{-1}$, $\lambda_C= \lambda_1^{-1}$, and
$\lambda_D= \lambda_2$. {\bf Case } $2$:
 $\lambda_A=\lambda_1$, $\lambda_B=\lambda_2$, $\lambda_C=
\lambda_2^{-1}$, and $\lambda_D= \lambda_1^{-1}$. $\P$} \label{fig2}
\end{figure}
We fix the branch of $\phi(z)$ by requiring that $ \phi(\infty)>0$.
We use $*$ to denote complex conjugation, and $\Xi$ is the unit
circle shown in Fig.~\ref{fig2}. The points $\lambda_1$ and
$\lambda_2$ are defined differently for the different values of
$\gamma$ and $h$.
\vfill\eject
{\Large There are   three different cases:}

 {\bf Case $1$a  is defined by inequality $4{(1-\gamma^2)}<h^2< 4$ }.

It describes moderate magnetic field.

 {\bf Case $2$ is defined by $h> 2$}. This is strong magnetic field.

 In both cases $\lambda_1$ and $\lambda_2$ are real and
given by the formulae
\begin{eqnarray}
\lambda_1=\frac{h-\sqrt{h^2-4(1-\gamma^2)}}{2(1+\gamma)},\quad
\lambda_2=\frac{1+\gamma}{1-\gamma} \lambda_1.\label{ldef1}
\end{eqnarray}

 {\bf Case $1$b is defined by $h^2<4(1-\gamma^2)$}.

  It is  weak magnetic field, including
 zero magnetic field.

  Both $\lambda_1$
and $\lambda_2$ are complex and given by the equations
\begin{eqnarray}
\lambda_1=\frac{h-\mathrm{i} \sqrt{4(1-\gamma^2)-h^2}}{2
(1+\gamma)},\quad
 \lambda_2=1/\lambda_1^*.\label{ldef2}
\end{eqnarray}
Note that in Case $1$ the poles of the function $\phi(z)$
(Eq.~\ref{defph}) coincide with the points $\lambda_{A}$ and
$\lambda_{B}$, while in  Case $2$ they coincide with the points
$\lambda_{A}$ and $\lambda_{C}$.

  We are going now to formulate our main results. To this end we need
  to introduce the Jacobi theta-function,
   \begin{eqnarray}
  \theta_3(s)=\sum_{n=-\infty}^{\infty} e^{\pi i \tau n^2+2\pi i s
  n}.\label{jacMay}
  \end{eqnarray}
  We remind the following characteristic properties of this function (see e.g. \cite{ww}):
  \begin{eqnarray}
  \theta_3(-s)=\theta_3(s),\quad \theta_3(s+1)=\theta_3(s) \label{theta1May}\\
  \theta_3(s+\tau)=e^{-\pi i \tau-2\pi i s} \theta_3(s) \label{theta2May}\\
  \theta_3\left(n+m\tau+\frac{1}{2}+\frac{\tau}{2}\right)=0,\quad
  n,m\in {\Bbb Z} \label{thetazerosMay}
  \end{eqnarray}
The modulus parameter $\tau$, in our case, is determined by the physical quantities
$h$ and $\gamma$ according to the
following {equation}\label{tauMay06}
\begin{equation}
  \tau =  i\tau_{0}, \quad  \tau_0= I(k')/I(k)
\end{equation}
where  $I(k)$ denotes the complete elliptic integral of the first kind,
$$
I(k) = \int_{0}^{1}\frac{dx}{\sqrt{(1-x^2)(1 - k^{2}x^{2})}} \qquad \qquad \diamond
$$
also $k'=\sqrt{1-k^2}$, and
  \begin{eqnarray}
   k= \left \{ \begin {array} {c} \sqrt{(h/2)^2+\gamma^2-1}\; /\; \gamma ,
  \;\;\;\mbox{Case 1a} \\ [0.3cm]
  \sqrt{1 -\gamma^2 - (h/2)^2}\; / \;\sqrt{1-(h/2)^2},
\;\;\; \mbox{Case 1b}\\ [0.3cm]
         \gamma\; / \;\sqrt{(h/2)^2+\gamma^2-1} ,\;\;\; \mbox{Case 2}
 \end{array}
  \right.
    \label{modMay}
  \end{eqnarray}
 Define also the function
 \begin{equation}\label{betadefMay}
 \beta(\lambda) = \frac{1}{2\pi i}\ln \frac{\lambda+1}{\lambda-1},
 \end{equation}
 and the infinite sequence of real numbers,
\begin{equation}\label{zerosMay}
\lambda_{m} =
  \tanh \left(m + \frac{1-\sigma}{2}\right)\pi \tau_{0}, \quad m \geq 0,
\end{equation}
where $\sigma = 1$ in Case 1 and $\sigma = 0$ in Case 2. Observe
that
\begin{itemize}
\item $0 < \lambda_{m} < 1, \quad \lambda_{m} \to 1, \quad m \to \infty$
\item in view of Eq.~(\ref{thetazerosMay}), the points $\pm \lambda_{m}$
are zeros of  $\theta_3\left(\beta(\lambda)+\frac{\sigma
\tau}{2}\right)$.
\end{itemize}

\medskip

{\bf Theorem 1.} {\it Let  $\Omega$ be the complex $\lambda$ - plane without
arbitrary fixed neighborhoods of the points $\lambda = \infty$,
 $\lambda = \pm 1$ and $\lambda = \pm \lambda_{m}, \quad m =0, 1, ...$ .
Then the Toeplitz determinant $D_L(\lambda)$ admits the following
asymptotic representation, which is uniform in $\lambda \in \Omega$.
\begin{equation}\label{DLasMay}
 \frac{d}{d\lambda}\ln D_L(\lambda)
= -\frac{2\lambda}{1-\lambda^{2}}L
+\frac{d}{d\lambda}
  \ln \left[ \theta_3\left(
  \beta(\lambda)+\frac{\sigma \tau}{2}\right)
  \theta_3\left(\beta(\lambda)-\frac{\sigma \tau}{2}\right)
  \right]
  \end{equation}
$$
+ O\left(\rho^{-L}\right), \quad L \to \infty.
$$
Here $\rho$ is any positive number satisfying the inequality,
$$
1 < \rho < |\lambda_{C}|.
$$}

This theorem  shows that  in the large $L$ limit,
the points $\pm\lambda_{m}$ (\ref{zerosMay}) are double zeros  of the
$D_L  (\lambda) $. More precisely, we see that in the
 large $L$ limit the eigenvalues $\nu_{2m}$ and  $\nu_{2m+1}$ from
  (\ref{eaap1}), (\ref{vd})  merge to $\lambda_{m}$:
\begin{equation}\label{doubleMay}
 \nu_{2m}, \nu_{2m+1} \to \lambda_{m},
\end{equation}
which in turn implies our {\bf main result}:

{\it The limiting entropy, $S(\rho_{A})$, of the subsystem
can be identified with the infinite
convergent series,
\begin{equation}\label{3333May}
 \diamondsuit  \quad \qquad  S(\rho_A) = \sum_{m=-\infty}^{\infty} H(\lambda_m) =\sum_{m=-\infty}^{\infty}
  (1+\lambda_{m})\ln \frac{2}{1+\lambda_{m}}  \qquad  \diamondsuit
  \end{equation}}

It is worth mentioning that relation (\ref{doubleMay}) also
indicates the degeneracy of the spectrum of the matrix ${\bf B}_{L}$
and  an appearance of an {\bf extra symmetry} in the large $L$
limit.

Observe that equation (\ref{eaa})
  can be also rewritten as
   \begin{eqnarray}
  S_{L}(\rho_A)=\lim_{\epsilon \to 0^+} \frac{1}{4\pi \mathrm{i}}
  \oint_{\Gamma'} \mathrm{d} \lambda\,  e(1+\epsilon, \lambda)
  \frac{\mathrm{d}}{\mathrm{d} \lambda} \ln
 \left( D_{L}(\lambda)(\lambda^{2} - 1)^{-L}\right)\;,\label{eaaMay0}
  \end{eqnarray}
because of the following identity
$$
 \lim_{\epsilon \to 0^+}
  \oint_{\Gamma'} \mathrm{d} \lambda\,  e(1+\epsilon, \lambda)
  \frac{\mathrm{d}}{\mathrm{d} \lambda} \ln
(\lambda^{2} - 1)^{-L} = L
\lim_{\epsilon \to 0^+}
  \oint_{\Gamma'} \mathrm{d} \lambda\,  e(1+\epsilon, \lambda)
  \frac{2\lambda}{1-\lambda^{2} }
$$
$$
=2\pi iL \lim_{\epsilon \to 0^+}
\left((2+\epsilon)\ln{\frac{2+\epsilon}{2}} + \epsilon \ln{\frac{\epsilon}{2}}\right)
= 0.
$$
In this paper, we  will  show that
  series (\ref{3333May}) coincides with
  the result of the following double limit procedure
   \begin{eqnarray}
  S(\rho_A)=\lim_{\epsilon \to 0^+}\left[\lim_{L\to \infty} \frac{1}{4\pi \mathrm{i}}
  \oint_{\Gamma'} \mathrm{d} \lambda\,  e(1+\epsilon, \lambda)
  \frac{\mathrm{d}}{\mathrm{d} \lambda} \ln
  \left(D_{L}(\lambda)(\lambda^{2} - 1)^{-L}\right)\right]\;,\label{eaaMay}
  \end{eqnarray}
  here the contour \mbox{$\Gamma'$} is depicted in Fig~$\ref{fig1}$.
  This result can be alternatively written down as the following integral,
  \begin{equation}\label{33May}
 \clubsuit  \qquad S(\rho_{A})
    = \frac{1}{2}\int_{1}^{\infty}\ln
  \left(\frac{\theta_{3}\left(\beta(\lambda) + \frac{\sigma \tau}{2}\right)
  \theta_{3}\left(\beta(\lambda) - \frac{\sigma \tau}{2}\right)}
  {\theta^{2}_{3}\left(\frac{\sigma \tau}{2}\right)}\right)\, d\lambda. \quad \spadesuit
 \end{equation}

We conjecture that in fact the limits in (\ref{eaaMay}) can be
interchanged. After the shorten version of this paper appeared in
quant-ph,  I.Peschel \cite{pes} simplified our expression for the
entropy for non-vanishing magnetic field   [Cases 1a and 2]. He used the approach of \cite{cardy}. He
showed that in these cases our formula (\ref{3333May}) is equivalent
to formula (4.33) of \cite{cardy}. Moreover, I. Peschel was able to
sum it up into the following expressions for the entropy.
\begin{eqnarray}\label{case1}
 S(\rho_A)=   \frac {1} {6} \left [\;\ln{ \left (\frac {k^2} {16 k'}\right )} +
\left (1-\frac {k^2} {2}\right )
         \frac {4 I(k) I(k')} {\pi} \right ] + \ln\;2 ,
    \end{eqnarray}
in {\bf Case 1a}, and
\begin{eqnarray}
   S(\rho_A)=  \frac {1} {12} \left [\;\ln{ \frac {16} {(k^2 k'^2)}} +
(k^2-k'^2)
         \frac {4 I(k) I(k')} {\pi} \right ],
   \end{eqnarray}
in {\bf Case 2 }. Here, $I(k)$ denotes the complete elliptic
integral of the first kind, $k'=\sqrt{1-k^2}$, and
\begin{eqnarray}
 k= \left \{ \begin {array} {c} \sqrt{(h/2)^2+\gamma^2-1}\; /\; \gamma
, \;\;\;\mbox{Case 1a} \\ [0.3cm]
       \gamma\; / \;\sqrt{(h/2)^2+\gamma^2-1} ,\;\;\; \mbox{Case 2}
\end{array} \right.
  \label{modMay2}
\end{eqnarray}
These describes cases of moderate and strong magnetic field.

We used the  equation
(\ref{3333May}) to calculate limiting entropy  for weak magnetic field.

In {\bf Case 1b} we  derived:
\begin{eqnarray}\label{case1b}
 S(\rho_A ) & =   \frac {1} {6} \left [\;\ln{ \left (\frac {k^2} {16 k'}\right )} +
\left (1-\frac {k^2} {2}\right )
         \frac {4 I(k) I(k')} {\pi} \right ] + \ln\;2 ,  \nonumber \\
 & k=\sqrt{\frac{1-h^2/4-\gamma^2}{1-h^2/4}} \qquad \qquad  \qquad \heartsuit  \nonumber \\
 \mbox{iff}  & 0\le h < 2\sqrt{1-\gamma^2}
    \end{eqnarray}
Note that this case  includes zero magnetic field.

These expressions helped us to study the range of variation of limting entropy.
Together with Dr. Franchini we found  that the
entropy  has a local minimum  at the
boundary of {\bf Cases 1a and 1b}:
$$ S(\rho)= \ln 2    \qquad  \mbox{at}  \qquad \left( \frac{h}{2}\right)^2+\gamma^2=1 \qquad  \qquad \star $$
At this boundary  the ground state is doubly degenerated, but the rest of energy levels are separated by a  gap.

Note that the  absolute minimum of asymptotic entropy  $\min S(\rho)=0$ is achieved at
infinite magnetic field corresponding to $k=0$ in Case 2 [the ground state is ferromagnetic]

\section{The Asymptotic of Block Toeplitz Determinants, Widom's Theorems.}

Our objective is the asymptotic calculation of the  block Toeplitz determinant
$D_L(\lambda)$  or, rather, its $\lambda$ -derivative $\frac{d}{d\lambda}\ln D_L(\lambda)$.
A general asymptotic representation of the determinant of a block
Toeplitz matrix, which generalizes the classical strong Szeg\"o theorem to
the block matrix case, was obtained by  H. Widom in \cite{widom} (see also
more recent work \cite{bottcher} and references therein). Here is Widom's result.

Let
$$
\varphi(z) = \sum_{k=-\infty}^{\infty}\varphi_{k}z^{k}, \quad |z| = 1,
$$
be an $p\times p$ matrix - valued function defined on the
unite circle, $|z| =1$, and satisfying the following conditions.
\begin{enumerate}
\item $||\varphi||
\equiv \sum_{k=-\infty}^{\infty}||\varphi_{k}|| +
\left\{\sum_{k=-\infty}^{\infty}|k|||\varphi_{k}||^2\right\}^{1/2} < \infty $
\item $\det \varphi(z) \neq 0, \quad \Delta|_{|z| = 1}\arg \det \varphi(z) =0$
\end{enumerate}
Consider the block Toeplitz determinant generated by $\varphi(z)$,
i.e. $D_{L}[\varphi] =
\det T_{L}[\varphi]$ where $T_{L}[\varphi] = \left(\varphi_{j-k}\right)$,
$j,k = 0, ..., L-1$.

{\bf Theorem 2.} (\cite{widom2}){\it Define{\footnote{ The
integration along the unit circle is always assumed to be done in
the positive, i.e. anti-clockwise direction.}}
\begin{equation}\label{widG}
G[\varphi] = \exp\left\{\frac{1}{2\pi i}\int_{|z| = 1}\ln \det \varphi(z)\frac{dz}{z}\right\}.
\end{equation}
Then the limit
\begin{equation}\label{widomtheor1}
E[\varphi] = {\lim_{L\to \infty}\frac{D_{L}[\varphi]}{\left(G[\varphi]\right)^{L}}}\,\,,
\end{equation}
exists. Moreover,  the following general formula
can be written for the quantity $E[\varphi]$,
\begin{equation}\label{widE}
E[\varphi] = \det \left(T_{\infty}[\varphi]T_{\infty}[\varphi^{-1}]\right).
\end{equation} }

This quite beautiful theorem is not  very efficient in concrete
applications. The remarkable fact though is that the proof
of theorem 2 is based on an auxiliary fact which was established in the
preceding work of Widom and which can be made an efficient
tool for a large class of symbols $\varphi$, including the
matrix function $\Phi(z)$ from (\ref{defphi})
which we are concerned with in this paper.

{\bf Theorem 3} (theorem 4.1 of \cite{widom}){\it Suppose that,
in addition to the conditions of theorem 2, the matrix symbol
 $\varphi^{-1}$ admits the Weiner-Hopf factorization,
\begin{equation}\label{widFact}
\varphi^{-1}(z) = u_{+}(z)u_{-}(z) = v_{-}(z)v_{+}(z),
\end{equation}
where the subscribes ``+''  and ``-'' indicate the analyticity inside
and outside of the unite circle, respectively. Suppose also that
$\varphi(z)$ can be included into a differentiable family,
$\lambda \to \varphi(z, \lambda)$. Then, $\ln E[\varphi]$
is a differentiable function of $\lambda $ and in fact,
\begin{equation}\label{widFact1}
\frac{d}{d\lambda}\ln E[\varphi]
=\frac{i}{2\pi}\int_{|z|=1}trace\left[
\left(u'_{+}(z)u_{-}(z) - v'_{-}(z)v_{+}(z)\right)
\frac{\partial \varphi(z,\lambda)}{\partial \lambda}\right]dz,
\end{equation}
where $(')$ means the derivative with respect to $z$.}

Let $\Omega$ be the set on the $\lambda$ - plane introduced in
theorem 1. In the next section we will show that for every $\lambda \in \Omega$,
the matrix-valued function $\Phi(z)$ admits the {\it explicit} Weiner-Hopf factorization,
and hence the second Widom theorem - theorem 3 above,  is applicable. Indeed, for the function $\Phi(z)$ this theorem can be specified as follows.

{\bf Theorem 4.} {\it Let $\lambda \in \Omega$ then the following
asymptotic representation for the logarithmic derivative of
the determinant $D_{L}(\lambda) = \det T_{L}[\Phi]$ takes place.
$$
\frac{d}{d\lambda}\ln
  D_L(\lambda) = -\frac{2\lambda}{1-\lambda^{2}}L \
$$
\begin{equation}\label{our}
+ \frac{1}{2\pi} \int_{|z| =
1}\mbox{trace}\, \Bigl[\left(U_{+}'(z)U_{+}^{-1}(z)
+V_{+}^{-1}(z)V_{+}'(z)\right)\Phi^{-1}(z)\Bigr]dz
\end{equation}
$$
+ r_{L}(\lambda),
$$
where the error term $r_{L}(\lambda)$ satisfies the uniform estimate,
\begin{equation}\label{errorour}
|r_{L}(\lambda)|\leq C\rho^{-L}, \quad
 \quad \lambda \in \Omega, \quad L \geq 1,
\end{equation}
and $\rho$ is any positive number such that $1 < \rho < |\lambda_{C}|$.
In (\ref{our}) the $2\times 2$ matrix-valued functions $U_{\pm}(z)$ and
$V_{\pm}(z)$  solve
the following  Weiner-Hopf factorization problem :
\begin{description}
\item {(i)}\quad $\Phi(z)=U_+(z)U_-(z)=V_-(z)V_+(z)
\quad z\in \Xi $ \item {(ii)} \quad $U_-(z)$ and $V_-(z)$ ($U_+(z)$ and
$V_+(z)$) are analytic outside (inside) the unit circle $\Xi$.
\item {(iii)} \quad $U_-(\infty)=V_-(\infty)=I$.
\end{description}}

{\bf Remark .} {\it The replacement of the factorization of the
inverse symbol by the factorization of the symbol itself has been made
for a purely technical reason.}

Asymptotic representation (\ref{our}) plays an important role  in our analysis
as, in fact, its starting point.  The truth of the matter is that we had derived
formula (\ref{our})
before we became aware of Widom's second theorem. In our original
derivation of (\ref{our}) and proof of theorem 4 we used
an alternative approach to Toeplitz determinants
suggested by P. Deift in \cite{deift}. It is based on the Riemann-Hilbert technique
of the theory of  ``integrable integral operators'', which was
 developed in \cite{iiks}, \cite{korepin} for evaluation of correlation functions of
  quantum
  completely integrable (exactly solvable) models
  {\footnote{In its turn, the approach of \cite{iiks} is based on
  the ideas of  \cite{jmms}. Several principal aspects of the
  integrable operator theory,
  especially the ones concerning with the integrable differential systems
  appearing in random matrix theory,
  have been developed in \cite{tw}. Some of the important elements of  modern theory of
  integrable operators were already implicitly present in the
  earlier work \cite{sakh}.}}. It turns out that, using the block matrix
  version of \cite{iiks} suggested in \cite{hi}, one can
  generalize Deift's scheme to the block Toeplitz matrices, which
  provides a rather simple derivation of equation (\ref{our}) together
  with the estimation of the error term indicated in theorem 4.

The  Riemann-Hilbert approach has already proved its usefulness
in the theory of  Toeplitz and Hankel determinants with scalar symbols (see
e.g. \cite{bdj}, \cite{bi}, \cite{dkmvz}, \cite{krasovsky}, \cite{dikz}).
Therefore, we   believe that the block-version of the
Riemann-Hilbert scheme is worthwhile to present. Indeed, although
in the relatively simple case of  smooth symbols
theorem 4, as it  turned out, is a direct (up to the error term estimation)
corollary of one of the old Widom's results, it is conceivable that in more
challenging  situations of the matrix symbols with singularities
the Riemann-Hilbert scheme might become very useful, as it has
been already the case for the scalar symbols.

Having all the above reasons in mind, we have decided to include in the
paper our original proof of theorem 4. We will do this in the
next section providing all the necessary facts concerning integrable
Fredholm operators and the Riemann-Hilbert method of their analysis.

The explicit factorization of the matrix $\Phi(z)$ will be performed
in section 4 and the final proof of our main result - theorem 1,
will be done in section 5.

\section{The Riemann-Hilbert approach to the block Toeplitz determinants.
An alternative proof of theorem 4.}

\subsection{The Fredholm Determinant Representation}

  Let $f_{j}(z)$ and $h_{j}(z)$, $j = 1, 2$, be $2\times 2$ matrix
  functions. We introduce the class of
  {\bf integrable operators} $K$  defined on
  $L_{2}(\Xi, {\Bbb C}^2)$ by the following equations (cf. \cite{hi}),
  \begin{equation}
  (K\, X)(z) = \oint_{\Xi}K(z,z')\, X(z')dz'\quad
  \textrm{for}\quad X\in L_{2},
  \end{equation}
  where
  $$
  K(z,z') =
  \frac{f^{T}(z)h(z')}{z-z'},\quad f(z) = \left(
  \begin{array}{c}
  f_{1}^{T}(z)\\
  f_{2}^{T}(z)
  \end{array}\right)
  $$
  \begin{equation}\label{intkernel}
  h(z) = \left( \begin{array}{c}
  h_{1}(z)\\
  h_{2}(z) \end{array}\right).
  \end{equation}
  Let $I_{2}$ denote the $2\times 2$ identity matrix. Put
  \begin{equation}
  f_{1}(z) = z^L I_{2}, \quad f_{2}(z) = I_{2} \label{f12}
  \end{equation}
  \begin{equation}\label{h12}
  h_{1}(z) = z^{-L}\frac{I_{2} - \Phi(z)}{2\pi i},\quad h_{2}(z) =
  -\frac{I_{2} - \Phi(z)}{2\pi i}.
  \end{equation}
  Then, essentially repeating the arguments of \cite{deift}, we have
  the following relation
  \begin{equation}\label{fredholm}
  D_L(\lambda) = \det (I - K),
  \end{equation}
where $I$ is the identity operator in $L_{2}(\Xi, {\Bbb C}^2)$.
The determinant in the l.h.s. is the
$2L\times 2L$ matrix determinant, while the (Fredholm) determinant in the r.h.s. is
taken in $L_{2}(\Xi, {\Bbb C}^2)$. This relation can be proved
by writing down the matrix representation  of the integral
operator $I - K$ in the basis $\{z^ke_{j}\}_{-\infty < k <
\infty,\, j =1,2}$, where $\{e_{1}, e_{2}\}$ is the canonical
basic in ${\Bbb C}^2$ (cf. \cite{deift}). The matrix elements of
operator $I-K$ can be defined by following relations:
\begin{eqnarray}
&&(I-K)z^k\, e_{\alpha}=\sum_{j=0}^{L-1}\sum_{\beta=1}^2
\Phi_{\beta,
\alpha}^{j-k} z^j\, e_{\beta}, \quad \qquad\qquad 0\le k< L, \quad \alpha=\{1,2\},\\
&&(I-K)z^k\, e_{\alpha}=z^k\, e_{\alpha}
+\sum_{j=0}^{L-1}\sum_{\beta=1}^2 \Phi_{\beta, \alpha}^{j-k} z^j\,
e_{\beta}, \quad k< 0 ~\textrm{or}~ k\ge L,\quad \alpha=\{1,2\}.
\end{eqnarray}
Here
$$
\Phi_{\beta, \alpha}^j = \frac{1}{2\pi
i}\int_{|z|=1}\Bigl(\Phi(z)\Bigr)_{\beta, \alpha} \frac{dz}
{z^{j+1}}
 \equiv  (\widetilde{\Pi}_j)_{\beta, \alpha}.
$$
With $\widetilde{\mathbf{B}}_{L}(\lambda)$  defined as $(\Phi_{\beta,
\alpha}^{j-k})$, $0\le j,k\le L-1$; $\alpha, \beta$ taking $1$
or $2$; $(j,\beta)$ numerating the rows and
$(k,\alpha)$ numerating the columns , equation (\ref{fredholm}) becomes obvious.

  So we represented $D_L(\lambda) $ as a Fredholm determinant  of
  the integral operator $K$.
  Define the resolvent operator $R$ by
$$
(I-K)(I+R)=I\quad \Leftrightarrow\quad
R = (I-K)^{-1}K.
$$
Then, we have the general equation \cite{iiks}, \cite{hi},
\begin{equation}\label{resolvent}
R(z,z') = \frac{F^{T}(z)H(z')}{z-z'},
\end{equation}
where
$$
F^{T} = (I-K)^{-1}f^{T}, \quad \mbox{and}\quad
H = h(I-K)^{-1}
$$
where in the first equation $(I-K)^{-1}$ is understood as acting to
the right, while in the second equation it acts to the left.

 \subsection{The Riemann-Hilbert Problem}

One of the main ingredients
of the theory of integrable operators is the following
Riemann-Hilbert representation for
the functions $F(z)$ and $H(z)$.
\begin{equation}\label{RHF}
F(z) = Y_{+}(z)f(z), \quad z\in \Xi,
\end{equation}
\begin{equation}\label{RHH}
H(z) = (Y^{T}_{+})^{-1}(z)h(z), \quad z\in \Xi,
\end{equation}
where the $4\times 4$ matrix function $Y(z)$ is the (unique)
solution of the following Riemann-Hilbert problem:
\begin{enumerate}
\item $Y(z)$ is analytic outside of the circle $\Xi$. \item
$Y(\infty) = I_{4}$, where $I_{4}$ denote the $4\times 4$ identity
matrix. \item $Y_{-}(z) = Y_{+}(z)J(z), \quad z\in \Xi$ where
$Y_{+}(z)$ ($Y_{-}(z)$) denote the left (right) boundary value of
$Y(z)$ on $\Xi$ (note, ``+'' means: from inside of the unite
circle!). The $4\times 4$ jump matrix $J(z)$ is defined by the
equations,
$$
J(z) = I_{4} + 2\pi if(z)h^{T}(z)
$$
\begin{equation}\label{J} = \left( \begin{array}{cc}
2I_{2} - \Phi^{T}(z)  & -z^{L}(I_{2} - \Phi^{T}(z))\\
z^{-L}(I_{2} - \Phi^{T}(z))& \Phi^{T}(z)\end{array}\right)
\end{equation}
\end{enumerate}
It is also worth noticing the integral formulae,
\begin{equation}\label{Yint}
Y(z) = I_{4} - \int_{\Xi}\frac{F(z')h^{T}(z')}{z' - z}dz',
\end{equation}
and
\begin{equation}\label{Yinvint}
Y^{-1}(z) = I_{4} + \int_{\Xi}\frac{f(z')H^{T}(z')}{z' - z}dz'.
\end{equation}
We are going now
to relate the $\lambda$ - derivative of $\det(I-K)$ and the resolvent.
To this end we first notice that
$$
\frac{d}{d\lambda}\Phi(z) = iI_{2}.
$$
Therefore, we have that,
$$
\frac{d}{d\lambda}K(z,z') = -\frac{1}{2\pi }\frac{z^{L}(z')^{-L}
-1}{z-z'}\,I_{2} \equiv -iK(z, z')(I_{2} - \Phi(z'))^{-1} .
$$
\mbox{From} this it follows that
$$
\left[(I - K)^{-1}\frac{d}{d\lambda}K\right](z,z') =
-iR(z,z')(I_{2} - \Phi(z'))^{-1},
$$
and hence
$$
\frac{d}{d\lambda}\ln \det T_{L}\{\phi \} = -\mbox{Trace}\,
(I-K)^{-1}\frac{d}{d\lambda}K
$$
\begin{equation}\label{trace1}
= i\int_{\Xi}\mbox{trace}\,\left(R(z,z)(I_{2} -
\Phi(z))^{-1}\right)dz,
\end{equation}
where ``Trace'' means the trace taking in the space
$L_{2}(\Xi, {\Bbb C}^2)$,
while ``trace'' is the $2\times 2$ matrix trace.

\mbox{From} (\ref{resolvent}) we conclude that
$$
R(z,z) = \frac{dF^{T}(z)}{dz}H(z),
$$
which in turn implies the equation
\begin{equation}\label{tau}
\frac{d}{d\lambda}\ln \det T_{L}\{\phi \} =
i\int_{\Xi}\mbox{trace}\,\left(\frac{dF^{T}(z)}{dz}H(z) (I_{2}
- \Phi(z))^{-1}\right)dz.
\end{equation}

\subsection{The Asymptotic Solution of the Riemann-Hilbert Problem.}

Equation (\ref{tau}), together with (\ref{RHF}) and (\ref{RHH}) it
reduces the question to the asymptotic analysis of the solution
$Y(z)$ of the Riemann-Hilbert problem (1 - 3).  For the latter we
shall follow \cite{deift}.

The basic observation is that the jump matrix $J(z)$ admits the
following algebraic factorization,
\begin{equation}\label{Jfactor}
J(z) = M(z)J_{0}(z)N(z),
\end{equation}.
where
\begin{equation}\label{Mdef}
M(z) =
\left( \begin{array}{cc}
I_{2}   & z^{L}\left(I_{2} - (\Phi^{T})^{-1}(z)\right)\\
0_{2}& I_{2}\end{array}\right),
\end{equation}
\begin{equation}\label{Ndef}
N(z) =
\left( \begin{array}{cc}
I_{2} &  0_{2}\\
- z^{-L}\left(I_{2} - (\Phi^{T})^{-1}(z)\right)&
I_{2}\end{array}\right),
\end{equation}
and
\begin{equation}\label{J0def}
J_{0}(z)
=\left( \begin{array}{cc}
(\Phi^{T})^{-1}(z) &  0_{2}\\
0_{2} & \Phi^{T}(z)\end{array}\right).
\end{equation}

Choose now a small $\epsilon$ and  define the matrix function $X(z)$
according to the equations:
\begin{equation}\label{Xdef1}
X(z) = Y(z) \quad \mbox{if}\quad |z| > 1 + \epsilon,
\quad \mbox{or}\quad |z| < 1 -\epsilon,
\end{equation}
\begin{equation}\label{Xdef2}
X(z) = Y(z)M(z) \quad \mbox{if}\quad 1-\epsilon < |z| < 1,
\end{equation}
\begin{equation}\label{Xdef3}
X(z) = Y(z)N^{-1}(z) \quad \mbox{if}\quad 1 < |z| < 1 + \epsilon.
\end{equation}
The new function has a jump accross the unit circle $\Xi$ with
the jump matrix $J_{0}(z)$ and two more jumps - accross the
circles,
$$
\Xi_{1}: |z| = 1-\epsilon,\quad \mbox{jump matrix}\, \, M(z)
$$
and
$$
\Xi_{2}: |z| = 1+\epsilon, \quad \mbox{jump matrix}\,\, N(z).
$$
In other words, the original Rimeann-Hilbert
problem (1 - 3)  is equivalent to the problem
\begin{description}
\item{1$^0$.} $X(z)$ is analytic outside of the contour $\Gamma \equiv \Xi \cup \Xi_{1}
\cup \Xi_{2}$.
\item{2$^0$.} $X(\infty) = I_{4}$, where $I_{4}$ denote the $4\times 4$ identity
matrix.
\item{3$^0$.} The jumps of the function $X(z)$ across the contour $\Gamma$ are
given by the equations
\begin{itemize}
\item  $X_{-}(z) = X_{+}(z)M(z), \quad z\in \Xi_{1}$
\item  $X_{-}(z) = X_{+}(z)N(z), \quad z\in \Xi_{2}$
\item  $X_{-}(z) = X_{+}(z)J_{0}(z), \quad z\in\Xi$
\end{itemize}
where the jump matrices $M(z)$,  $N(z)$, and $J_{0}(z)$ are
defined in (\ref{Mdef}), (\ref{Ndef}), and (\ref{J0def}), respectively
and each circle is oriented counterclockwise.
\end{description}
Observe that the differences of the jump matrices on $\Xi_{1}$
and $\Xi_{2}$ from the identity are exponentially small as $n
\to \infty$. This means one can expect the following asymptotic relation for
$X(z)$,
\begin{equation}\label{asymp1}
X(z) \sim X^{0}(z),
\end{equation}
where the independent on $L$ function $X^{0}(z)$ solves the Riemann-Hilbert problem which is
the same as the $Y$ - problem but with the jump matrix $J_{0}(z)$
instead of $J(z)$.

We notice that the function $X^{0}(z)$ can be found explicitly in terms of
the $2\times 2$ matrix-valued functions $U_{\pm}(z)$ and
$V_{\pm}(z)$  solving
the following  Weiner-Hopf factorization problem :
\begin{description}
\item {(i)}\quad $\Phi(z)=U_+(z)U_-(z)=V_-(z)V_+(z) ,
\quad z\in \Xi $ \item {(ii)} \quad $U_-(z)$ and $V_-(z)$ ($U_+(z)$ and
$V_+(z)$) are analytic outside (inside) the unit circle $\Xi$.
\item {(iii)} \quad $U_-(\infty)=V_-(\infty)=I$.
\end{description}
Indeed we have  that
\begin{equation}\label{X0def1}
X^{0}(z)
=\left( \begin{array}{cc}
U_{+}^{T}(z) &  0_{2}\\
0_{2} & (V_{+}^{T})^{-1}(z)\end{array}\right), \quad \mbox{if}
\quad |z| < 1,
\end{equation}
and
\begin{equation}\label{X0def2}
X^{0}(z)
=\left( \begin{array}{cc}
(U_{-}^{T})^{-1}(z) &  0_{2}\\
0_{2} & V_{-}^{T}(z)\end{array}\right), \quad \mbox{if}
\quad |z| > 1.
\end{equation}

In the next section, we will show that for all $\lambda \in \Omega$
the above Weiner-Hopf factorization   can be  found explicitely,
in terms of the elliptic theta-functions. In this section
though we won't need the explicit formulae for $U_{\pm}(z)$ and $V_{\pm}(z)$.
The only information we will need in this section is the existence
of $U_{\pm}(z)$ and $V_{\pm}(z)$, and the uniform estimate
\begin{equation}\label{uvest}
|U_{\pm}(z)|,\quad |V_{\pm}(z)| < C, \quad \forall z, \quad \forall \lambda \in \Omega,
\end{equation}
which is the direct consequence of the explicit formulae (\ref{u-d1}), (\ref{u-d}),
and (\ref{uv1}), (\ref{uv2}) of section 4.

We are now ready to formulate and prove the rigorous version of the
formal asymptotic relation (\ref{asymp1}).

{\bf Theorem 5.} {\it Let $\Omega$ be the set introduced in theorem 1,
i.e. the complex $\lambda$-plane without arbitrary fixed neighborhoods
of the points $\lambda = \infty$, $\lambda = \pm 1$, and $\lambda = \pm\lambda_{m}$,
$m = 0, 1, ...$, where $\lambda_{m}$ are defined in (\ref{zeros}) .
Then, for sufficiently large $L$ and all $\lambda \in \Omega$,
the Riemann-Hilbert problem (1$^0$ - 3$^0$) has the unique solution $X(z)$
which satisfies the following  uniform estimate,
\begin{equation}\label{theorem5}
\left|X(z)\left(X^{0}(z)\right)^{-1} - I_{4}\right| \leq C\frac{\rho^{-L}}{1+|z|},\quad
z\in {\Bbb C}, \quad\lambda \in \Omega, \quad L \geq L_{0},
\end{equation}
where $\rho$ is any positive number such that $1 < \rho < |\lambda_{C}|$,
and the block diagonal matrix-valued function $X^{0}(z)$ is defined
by the equations (\ref{X0def1}) and (\ref{X0def2}).}

\noindent
{\bf Proof.} Put
$$
R(z): = X(z)\left(X^{0}(z)\right)^{-1}.
$$
Observe that since $\det J_{0}(z) \equiv 1$, the function
$d(z)=\det X^{0}(z)$ solves the following scalar Riemann-Hilbert problem:
\begin{itemize}
\item $d(z)$ is analytic outside the unit circle $\Xi$
\item $d_{-}(z) = d_{+}(z)$, $z \in \Xi$
\item $d(\infty) = 1$
\end{itemize}
By virtue of the Liouville theorem, this implies that $d(z)=\det X^{0}(z) \equiv 1$,
and hence the matrix ratio $R(z)$ is well defined. In fact,
in terms of the matrix function $R(z)$ the Riemann-Hilbert
problem (1$^0$ - 3$^0$) can be rewritten as follows .
\begin{description}
\item{1$^{00}$.} $R(z)$ is analytic outside of the contour $\Gamma_{0}
 \equiv \Xi_{1}
\cup \Xi_{2}$ (no jump across the unite circle $\Xi$!).
\item{2$^{00}$.} $R(\infty) = I_{4}$, where $I_{4}$ denote the $4\times 4$ identity
matrix.
\item{3$^{00}$.} The jumps of the function $R(z)$ across the contour $\Gamma_{0}$ are
given by the equations
\begin{itemize}
\item  $R_{-}(z) = R_{+}(z)\hat{M}(z), \quad z\in \Xi_{1}$
\item  $R_{-}(z) = R_{+}(z)\hat{N}(z), \quad z\in \Xi_{2}$
\end{itemize}
where the jump matrices $\hat{M}(z)$ and $\hat{N}(z)$ are defined
by the formulae
\begin{equation}\label{MNhat}
\hat{M}(z) = X^{0}(z)M(z)[X^{0}(z)]^{-1},\quad
\hat{N}(z) = X^{0}(z)N(z)[X^{0}(z)]^{-1}.
\end{equation}
\end{description}
Let us denote $G_{0}(z)$ the jump matrix of the function $R(z)$,
i.e.
 \begin{eqnarray}
   G_{0}(z)= \left \{ \begin {array} {c}
\hat{M}(z), \quad z\in \Xi_{1} \\ [0.3cm]
\hat{N}(z), \quad z\in \Xi_{2}\end{array}\right.\label{G0def}
  \end{eqnarray}
Then, taking into account the estimate (\ref{uvest}) and the
expressions for matrices $M(z)$ and $N(z)$ we immediately
arrive at the estimate,
\begin{equation}\label{Gest1}
|I_{4} - G_{0}(z)| \leq C\rho^{-L},\quad 1 < \rho < |\lambda_{C}|,
\quad z \in \Gamma_{0}, \quad \lambda \in \Omega, \quad L \geq 1.
\end{equation}
This in turn implies the $L_{2}\cap L_{\infty}$ - norm
estimate:
\begin{equation}\label{Gest2}
\|I_{4} - G_{0}\|_{L_{2}(\Gamma_{0})\cap L_{\infty}(\Gamma_{0})}\leq C\rho^{-L},
\end{equation}
which, by standard arguments  based on the analysis of the
relevant singular integral equation (see e.g. \cite{dkmvz}
or  Appendix D of \cite{bi}),
yields the estimate (\ref{theorem5}).
For the reader's convenience we will now present those arguments.

According to the general theory of Riemann-Hilbert
problems (see e.g. \cite{CG} or \cite{BDT})
the solution $R(z)$ of the Riemann-Hilbert problem (1$^{00}$ - 3$^{00}$)
is given by the integral representation,
\begin{equation}\label{RHMay1}
R(z) = I + {1\over 2\pi i}\int _{\Gamma_{0}}r(z') \left (I_{4}- G_{0}(z')
\right ){dz' \over {z' - z}},\quad z\not\in\Gamma_{0},
\end{equation}
where $r(z)\equiv R_{+}(z)$ solves the singular integral equation
\begin{equation}\label{RHMay2}
r(z) = I + C_{+}[r \left (I_{4}- G_{0}\right )](z),
\quad z\in \Gamma_{0},
\end{equation}
and $C_{+}$ is the corresponding Cauchy operator:
$$
(C_{+}h)(z) = \lim_{  z'\to z, \, z'\in
(+)-\mbox{side}}\int_{\Gamma_{0}}{h(s)\over {s - z'}}\,
{ds \over 2\pi i}.
$$
We note that in the case under consideration this statement
is an (almost) immediate consequence of the Cauchy theorem
applied to the Cauchy integral,
$$
{1\over 2\pi i}\int _{\Gamma_{0}}R_{+}(z') \left (I_{4}- G_{0}(z')
\right ){dz' \over {z' - z}}
$$
in conjunction with the jump relations 3$^{00}$.

Put
$$
r_{0}(z) = r(z) - I_{4}.
$$
Then equation (\ref{RHMay2}) can be rewritten as an equation in $L_{2}(\Gamma_{0})$,
\begin{equation}\label{RHMay3}
(1 - {\Bbb K})r_{0} = f,
\end{equation}
where the function $f(z)$ and the operator ${\Bbb K}$ are
defined by the formulae,
$$
f = C_{+}(I - G_{0}),
$$
$$
{\Bbb K}:h \mapsto C_{+}[h \left (I_{4}- G_{0}\right )],\quad h \in
L_{2}(\Gamma_{0}).
$$
The $L_{2}$-boundness of the operator $C_{+}$ (see e.g. \cite{lis})
and the estimate (\ref{Gest2}) imply that
$$
||{\Bbb K}||_{L_{2}(\Gamma_{0}) \to L_{2}(\Gamma_{0})} \leq
||C_{+}||_{L_{2}(\Gamma_{0}) \to L_{2}(\Gamma_{0})}
||I_{4}-G_{0}||_{L_{\infty}(\Gamma_{0})}
= O(\rho^{-L}),
$$
and
$$
||f||_{L_{2}(\Gamma_{0})} \leq ||C_{+}||_{L_{2}(\Gamma_{0}) \to L_{2}(\Gamma_{0})}
||I_{4}-G_{0}||_{L_{2}(\Gamma_{0})} = O(\rho^{-L}).
$$
Therefore, equation  (\ref{RHMay3}) is uniquely solvable in $L_{2}(\Gamma_{0})$ for
sufficiently large $L$, and its solution satisfies the estimate,
$$
||r_{0} ||_{L_{2}(\Gamma_{0})} = O(\rho^{-L}),
$$
or, in terms of $r(z)$,
$$
||I_{4}-r ||_{L_{2}(\Gamma_{0})} = O(\rho^{-L}).
$$
This equation together with (\ref{RHMay1}) allows us
to estimate $R(z)$:
$$
|R(z) - I_{4}| \leq C || I_{4} - G_{0}||_{L_{\infty}(\Gamma_{0})}
\frac{1}{\mbox{dist}\,\left(z;\Gamma_{0}\right)}
+ C||I_{4}-G_{0}||_{L_{2}(\Gamma_{0})}||I_{4}-r ||_{L_{2}(\Gamma_{0})}
\frac{1}{\mbox{dist}\,\left(z;\Gamma_{0}\right)}
$$
$$
\leq C\frac{\rho^{-L}}{\mbox{dist}\,\left(z;\Gamma_{0}\right)}.
$$
This leads to the announced estimate (\ref{theorem5}) for all $z$ outside of some neighborhood
of the contour $\Gamma_{0}$. Since of the flexibility in the choice
of the circles $\Xi_{1}$ and $\Xi_{2}$ (i.e. the flexibility in the choice
of $\epsilon$ in (\ref{Xdef1} - \ref{Xdef3})), the estimate is extended to the
neighborhood of the contour $\Gamma_{0}$.  This completes the proof of the theorem.

\subsection{The Proof of Theorem 4.}

Let us observe that in terms of the function $X(z)$
the equations (\ref{RHF}) and (\ref{RHH}), which
determine the functions $F(z)$ and $H(z)$, can be
transformed as follows.
$$
F(z) = X_{+}(z)M^{-1}(z)f(z)
$$
$$
= X_{+}(z)\left( \begin{array}{cc}
I_{2} & - z^{L}\left(I_{2} - (\Phi^{T})^{-1}(z)\right)\\
0_{2}&I_{2}\end{array}\right)
\left( \begin{array}{c}
z^{L}I_{2}\\
I_{2}\end{array}\right) = X_{+}(z)
\left( \begin{array}{c}
z^{L}(\Phi^{T})^{-1}(z)\\
I_{2}\end{array}\right),
$$
and, similarly,
$$
H(z)\left(I_{2} - \Phi(z)\right)^{-1} =
(X^{T}_{+})^{-1}(z)M^{T}(z)h(z)\left(I_{2} - \Phi(z)\right)^{-1}
$$
$$
= \frac{1}{2\pi i}(X^{T}_{+})^{-1}(z)\left( \begin{array}{cc}
I_{2} & 0_{2} \\
z^{L}\left(I_{2} - \Phi^{-1}(z)\right)&I_{2}\end{array}\right)
\left( \begin{array}{c}
z^{-L}I_{2}\\
-I_{2}\end{array}\right) = \frac{1}{2\pi i}(X^{T}_{+})^{-1}(z)
\left( \begin{array}{c}
z^{-L}I_{2}\\
-\Phi^{-1}(z)\end{array}\right).
$$
This implies that
$$
\mathrm{tr}\,\left(\frac{dF^{T}(z)}{dz}H(z) (I_{2}
- \Phi(z))^{-1}\right)
=\frac{1}{2\pi i}\frac{L}{z}\mathrm{tr}\,\Phi^{-1}(z)
+ \frac{1}{2\pi i}\frac{d}{dz}\mathrm{tr}\,\Phi^{-1}(z)
$$

$$
+\frac{1}{2\pi i}\mathrm{tr}\,\left[
\frac{dX^{T}_{+}(z)}{dz}(X^{T}_{+})^{-1}(z)
\left( \begin{array}{cc}
\Phi^{-1}(z) & z^{-L}I_{2} \\
-z^{L}\Phi^{-2}(z) & -\Phi^{-1}(z)\end{array}\right)\right],
$$
and hence, our basic relation for the Toeplitz
determinant (\ref{tau}) can be re-written as
$$
\frac{d}{d\lambda}\ln
  D_L(\lambda) = \frac{L}{2\pi} \int_{\Xi}\mathrm{tr}\,
\Phi^{-1}(z)\frac{dz}{z}
$$

\begin{equation}\label{taumarch24}
+ \frac{1}{2\pi} \int_{\Xi}\mathrm{tr}\,\left[
\frac{dX^{T}_{+}(z)}{dz}(X^{T}_{+})^{-1}(z)
\left( \begin{array}{cc}
\Phi^{-1}(z) & z^{-L}I_{2} \\
-z^{L}\Phi^{-2}(z) & -\Phi^{-1}(z)\end{array}\right)\right]dz.
\end{equation}

The second term in the r.h.s. of equation (\ref{taumarch24})
can be split into the following three integrals,
$$
\frac{1}{2\pi} \int_{\Xi}\mathrm{tr}\,\left[
\frac{dX^{T}_{+}(z)}{dz}(X^{T}_{+})^{-1}(z)
\left( \begin{array}{cc}
\Phi^{-1}(z) & z^{-L}I_{2} \\
-z^{L}\Phi^{-2}(z) & -\Phi^{-1}(z)\end{array}\right)\right]dz
$$
$$
=\frac{1}{2\pi} \int_{\Xi}\mathrm{tr}\,\left[
\frac{dX^{T}_{+}(z)}{dz}(X^{T}_{+})^{-1}(z)
\left( \begin{array}{cc}
\Phi^{-1}(z) & 0_{2} \\
0_{2}& -\Phi^{-1}(z)\end{array}\right)\right]dz
$$
$$
+\frac{1}{2\pi} \int_{\Xi}\mathrm{tr}\,\left[
\frac{dX^{T}_{+}(z)}{dz}(X^{T}_{+})^{-1}(z)
\left( \begin{array}{cc}
0_{2} & 0_{2} \\
-z^{L}\Phi^{-2}(z) & 0_{2}\end{array}\right)\right]dz
$$
$$
+\frac{1}{2\pi} \int_{\Xi}\mathrm{tr}\,\left[
\frac{dX^{T}_{+}(z)}{dz}(X^{T}_{+})^{-1}(z)
\left( \begin{array}{cc}
0_{2} & z^{-L}I_{2} \\
0_{2}& 0_{2}\end{array}\right)\right]dz
$$
\begin{equation}\label{march241}
\equiv {\Bbb T}_{1} + {\Bbb T}_{2} + {\Bbb T}_{3}.
\end{equation}
The integral ${\Bbb T}_{2}$ can be replaced by the integral
over the contour $\Xi_{1}$,
\begin{equation}\label{T2}
{\Bbb T}_{2} =
\frac{1}{2\pi} \int_{\Xi_{1}}\mathrm{tr}\,\left[
\frac{dX^{T}_{-}(z)}{dz}(X^{T}_{-})^{-1}(z)
\left( \begin{array}{cc}
0_{2} & 0_{2} \\
-z^{L}\Phi^{-2}(z) & 0_{2}\end{array}\right)\right]dz,
\end{equation}
and hence estimated as
\begin{equation}\label{estT2}
|{\Bbb T}_{2}| \leq C\rho^{-L}, \quad
1 < \rho < |\lambda_{C}|, \quad \lambda \in \Omega, \quad L \geq L_{0}.
\end{equation}
(We notice that by virtue of theorem 5, the matrix function $X(z)$ is
uniformly bounded)
Using the jump relation,  $X_{-}(z) = X_{+}(z)J_{0}(z)$, the integral
${\Bbb T}_{3}$ can be rewritten as
$$
{\Bbb T}_{3}
= \frac{1}{2\pi} \int_{\Xi}\mathrm{tr}\,\left[
\frac{dX^{T}_{-}(z)}{dz}(X^{T}_{-})^{-1}(z)
\left( \begin{array}{cc}
0_{2} & z^{-L}\Phi^{-2}(z) \\
0_{2}& 0_{2}\end{array}\right)\right]dz,
$$
and, similarly to the integral ${\Bbb T}_{2}$, further replaced
by the integral over the contour $\Xi_{2}$,
\begin{equation}\label{T3}
{\Bbb T}_{3}
= \frac{1}{2\pi} \int_{\Xi_{2}}\mathrm{tr}\,\left[
\frac{dX^{T}_{+}(z)}{dz}(X^{T}_{+})^{-1}(z)
\left( \begin{array}{cc}
0_{2} & z^{-L}\Phi^{-2}(z) \\
0_{2}& 0_{2}\end{array}\right)\right]dz,
\end{equation}
which in turn yields the same estimate (\ref{estT2}) as in the case
of integral ${\Bbb T}_{2}$,
\begin{equation}\label{estT3}
|{\Bbb T}_{3}| \leq C\rho^{-L}, \quad
1 < \rho < |\lambda_{C}|, \quad \lambda \in \Omega, \quad L \geq 1.
\end{equation}
Finally we notice that, by virtue of theorem 5,  the matrix function
$X_{+}(z)$ in the integral ${\Bbb T}_{1}$ can be replaced, within  the same error
(\ref{estT3}),  by the
block diagonal matrix function $X^{0}(z)$ from equation
(\ref{X0def1}). Therefore, we derive from (\ref{taumarch24}) the following asymptotic
representation for the logarithmic derivative of $D_{L}(\lambda)$.
$$
\frac{d}{d\lambda}\ln
  D_L(\lambda) = \frac{L}{2\pi} \int_{\Xi}\mathrm{tr}\,
\Phi^{-1}(z)\frac{dz}{z}
$$
\begin{equation}\label{taumay24}
+ \frac{1}{2\pi} \int_{\Xi}\mathrm{tr}\,\left[
\frac{d(X^{0}_{+})^{T}(z)}{dz}((X^{0}_{+})^{T})^{-1}(z)
\left( \begin{array}{cc}
\Phi^{-1}(z) & 0_{2} \\
0_{2} & -\Phi^{-1}(z)\end{array}\right)\right]dz
+ r_{L}(\lambda)
\end{equation}
where the error term $r_{L}(\lambda)$ satisfies the estimates,
\begin{equation}\label{rest1}
|r_{L}(\lambda)|\leq C\rho^{-L}, \quad
1 < \rho < |\lambda_{C}|, \quad \lambda \in \Omega, \quad L \geq 1
\end{equation}
Taking into account formula (\ref{X0def1})
for $X^{0}(z)$ we can simplify (\ref{taumay24}) as follows,
$$
\frac{d}{d\lambda}\ln
  D_L(\lambda) = \frac{L}{2\pi} \int_{\Xi}\mathrm{tr}\,
\Phi^{-1}(z)\frac{dz}{z}
$$
\begin{equation}\label{widom1}
+ \frac{1}{2\pi} \int_{|z| =
1}\mbox{trace}\, \left[\left(U_{+}'(z)U_{+}^{-1}(z)
+V_{+}^{-1}(z)V_{+}'(z)\right)\Phi^{-1}(z)\right]dz
\end{equation}
$$
+ r_{L}(\lambda),
$$
or

$$
\frac{d}{d\lambda}\ln
  D_L(\lambda) = -\frac{2\lambda}{1-\lambda^{2}}L \
$$
\begin{equation}\label{widom2}
+ \frac{1}{2\pi} \int_{|z| =
1}\mbox{trace}\, \left[\left(U_{+}'(z)U_{+}^{-1}(z)
+V_{+}^{-1}(z)V_{+}'(z)\right)\Phi^{-1}(z)\right]dz
\end{equation}
$$
+ r_{L}(\lambda),
$$
where we have used the relation
$$
\mathrm{tr}\,\Phi^{-1}(z) = \frac{2i\lambda}{1-\lambda^2}\,.
$$
In the last formulae, the symbol $(')$ means the derivative with respect to $z$.
This completes the proof of theorem 4.

\section{Wiener-Hopf Factorization of Matrix $\Phi(z)$}

  By explicit calculation, one can find that
  \begin{eqnarray}
  (1-\lambda^2) \sigma_3 \Phi^{-1}(z) \sigma_3= \Phi(z),\quad
  \sigma_3= \left( \begin{array}{cc} 1 &
  0\\
  0& -1\end{array}\right).
  \end{eqnarray}
  Hence,
  \begin{eqnarray}
  V_-(z)&=&\sigma_3 U_-^{-1}(z)\sigma_3 \label{uv1}\\
  V_+(z)&=&\sigma_3 U_+^{-1}(z)\sigma_3(1-\lambda^2),\quad
  \lambda\neq \pm 1,\label{uv2}
  \end{eqnarray}
  and one only needs the explicit expressions for $U_{\pm}(z)$.

  Our observation is that, for all $\lambda$
  outside of a certain discrete subset of the interval
  $[-1, 1]$, the solution to the auxiliary Riemann-Hilbert problem
  (i-iii) exists; moreover, the functions
  $U_{\pm}(z)$ can be expressed in terms of the Jacobi
  theta-functions. Indeed, the auxiliary Riemann-Hilbert
  problem (i-iii) can be easily reduced to a type of the
  ``finite-gap'' Riemann-Hilbert problems which
  have already appeared in the analysis of the
  integrable statistical mechanics models
  (see \cite{diz}). Before we give detail
  expressions, let us first define some basic objects:
  \begin{eqnarray}
  w(z)&=&\sqrt{(z-\lambda_1)
  (z-\lambda_2)(z-\lambda_2^{-1})(z-\lambda_1^{-1})},\label{notion1}\\
  \beta(\lambda)&=&\frac{1}{2\pi i}\ln
  \frac{\lambda+1}{\lambda-1},\label{betadef}\end{eqnarray} where $w(z)$ is
  analytic on the domain ${\Bbb C} \backslash \, \{ J_1\cup J_2\}$
  shown in Fig.~\ref{fig2} and fixed by the condition:
  $w(z) \to z^2$ as

   $z \to \infty $.
  Next we define
  \begin{eqnarray}
  \tau=\frac{2}{c} \int_{\lambda_B}^{\lambda_C}\frac{\mathrm{d}
  z}{w(z)}, \quad
  c=2\int_{\lambda_A}^{\lambda_B}\frac{\mathrm{d}z}{w(z)},\label{important}\end{eqnarray}
  \begin{eqnarray}
  \delta=\frac{2}{c} \left(-\pi i-\int_{\lambda_A}^{\lambda_B}
  \frac{z \mathrm{d} z}{w(z) }\right),  \quad \omega(z)= \frac{1}{c}
  \int_{\lambda_A}^{z}\frac{\mathrm{d} z}{w(z)},
  \end{eqnarray}
  \begin{equation}
   \Delta(z)=\frac{1}{2}\int_{\lambda_A}^z
  \frac{z+\delta}{w(z)}\mathrm{d} z,\quad
  \kappa=\int_{\lambda_A}^{\infty}\mathrm{d} \omega(z),
  \end{equation}
  Points $\lambda_A,\lambda_B,\lambda_C, \lambda_D$
  and cuts $J_1$, $J_2$ and curves $\Sigma$ and $\Xi$ are shown in
  Fig.~\ref{fig2}. We shall also need:
  \begin{equation}
  \Delta_0=\lim_{z\to \infty}\left[\Delta(z)-\frac{1}{2} \ln
  (z-\lambda_1)\right].\label{notion2}
  \end{equation}
  Here, the contours of integration
  for $c$ and $\delta$ are taken along the
  left side of the cut $J_{1}$. The contour of
  integration for $\tau$ is the segment
  $[\lambda_{B}, \lambda_{C}]$. The contours of integration
  for $\kappa$ and in (\ref{notion2}) are taken along the line $\Sigma$ to the left
  from $\lambda_{A}$; also in (\ref{notion2}), $\arg(z-\lambda_{1}) = \pi$.
  The contours of integration in the
  integrals $\Delta(z)$ and $\omega(z)$ are
  taking according to the rule: The contour lies entirely in
  the domain $\Omega_{+}$ ($\Omega_{-}$) for $z$ belonging to $\Omega_{+}$
  ($\Omega_{-}$). It also worth noticing that, for the general reasons
(see e.g. \cite{fk}; see also Chapter 2 of \cite{bbeim}),
we always have that Re$i\tau < 0$. Moreover, as we will see later on,
the following inequality takes place
\begin{equation}\label{riemann}
 i\tau < 0.
\end{equation}
It also should be noted that $\omega(z)$ is an elliptic integral
of the first kind, normalized by the period condition
\begin{equation}\label{omegaA}
2\int_{\lambda_{A}}^{\lambda_{B}}\mathrm{d} \omega(z)=1.
\end{equation}
and $\Delta(z)$ is an elliptic integral of the third kind, normalized
by the conditions
\begin{equation}\label{deltaA}
2\int_{\lambda_{A}}^{\lambda_{B}} \mathrm{d}
\Delta(z)=-\pi i,
\end{equation}
and
\begin{equation}\label{deltainf}
\Delta(z)\sim
\frac{1}{2}\ln z \quad\textrm{for $z\to \infty$}.
\end{equation}

The rules formulated above define the integrals $\Delta(z)$ and $\omega(z)$
as the single-valued analytic functions in the domains $\Omega_{+}$
and $\Omega_{-}$. Let us denote $\Sigma_{A}$ and $\Sigma_{D}$ the parts of the
contour $\Sigma$ which lie to the left of $\lambda_{A}\equiv \lambda_{1}$ and
to the right of  $\lambda_{D}\equiv \lambda_{1}^{-1}$, respectively{\footnote{ In particular,
for the Cases 1a and 2, we have that $\Sigma_{A}$ and $\Sigma_{D}$
are the half lines $(-\infty, \lambda_{A})$ and $(\lambda_{D}, +\infty)$,
respectively.}}. Then, on the polygonal line $\Sigma$ the integrals
$\Delta(z)$ and $\omega(z)$ have jumps which can be described by the equations,
\begin{eqnarray}
&&\omega_+(z)-\omega_-(z)=0~\textrm{and}~\Delta_+(z)-\Delta_-(z)=0 \quad \textrm{for} ~z \in \Sigma_{A},\label{jinq0}\\
&&\omega_+(z)+\omega_-(z)=0~\textrm{and}~ \Delta_+(z)+\Delta_-(z)=0\quad \textrm{for} ~z\in (\lambda_{A}, \lambda_{B})\\
&&\omega_+(z)-\omega_-(z)=1~\textrm{and}~ \Delta_+(z)-\Delta_-(z)=-\pi i\quad \textrm{for} ~z\in (\lambda_{B}, \lambda_{C})\\
&&\omega_+(z)+\omega_-(z)=\tau~\textrm{and}~\Delta_+(z)+\Delta_-(z)=2\int_{\lambda_{B}}^{\lambda_{C}}\mathrm{d}
\Delta(z) =-2\pi i\int_{\lambda_1}^{\infty}\mathrm{d}\omega(z)\nonumber\\
 &&\quad \textrm{for} ~z\in (\lambda_{C}, \lambda_{D}) ~\textrm{and integration is taken over the ray $\Sigma_{A}$}\label{jinq1} \\ &&\omega_+(z)-\omega_-(z)=0~\textrm{and}~
\Delta_+(z)-\Delta_-(z)=-\pi i\quad \textrm{for}~z\in \Sigma_{D}\label{jinq2}
\end{eqnarray}
The only relation above which is not straightforward is the last equation in (\ref{jinq1}).
It easy follows, however, from the application of the classical Riemann bilinear relations
(see e.g. \cite{fk} or \cite{bbeim}) to the periods abelian (elliptic) integrals $\omega(z)$  and
$\Delta (z)$.

  Let
   \begin{eqnarray}
  \theta_3(s)=\sum_{n=-\infty}^{\infty} e^{\pi i \tau n^2+2\pi i s
  n}\label{jac}
  \end{eqnarray}
 be the third  Jacobi theta-function.
We remind again the basic properties of the function$ \theta_3(s)$ (see e.g. \cite{ww}):
  \begin{eqnarray}
  \theta_3(-s)=\theta_3(s),\quad \theta_3(s+1)=\theta_3(s) \label{theta1}\\
  \theta_3(s+\tau)=e^{-\pi i \tau-2\pi i s} \theta_3(s) \label{theta2}\\
  \theta_3\left(n+m\tau+\frac{1}{2}+\frac{\tau}{2}\right)=0,\quad
  n,m\in {\Bbb Z} \label{thetazeros}
  \end{eqnarray}
  We also introduce the $2\times 2$
  matrix valued function $\Theta(z)$ with the entries,
  \begin{eqnarray}&&\Theta_{11}(z)=(z-\lambda_1)^{-\frac{1}{2}}
  e^{\Delta(z)}\nonumber\\&&\quad \quad \quad  \times
  \frac{\theta_3\left(\omega(z)+\beta(\lambda)-\kappa +\frac{\sigma
  \tau}{2}\right)}{\theta_3\left(\omega(z) + \frac{\sigma \tau}{2}\right)}\nonumber\\
  &&\Theta_{12}(z)=-(z-\lambda_1)^{-\frac{1}{2}} e^{-\Delta(z)}\nonumber\\ &&\quad
  \quad \quad \times \frac{\theta_3\left(\omega(z)-\beta(\lambda)+\kappa -\frac{\sigma
  \tau}{2}\right)}{\theta_3\left(\omega(z) - \frac{\sigma
  \tau}{2}\right)}\nonumber\\&&\Theta_{21}(z)=-(z-\lambda_1)^{-\frac{1}{2}}
  e^{-\Delta(z)}\nonumber\\ &&\quad \quad \quad \times
  \frac{\theta_3\left(\omega(z)+\beta(\lambda) +\kappa -\frac{\sigma \tau}{2}\right)}
  {\theta_3\left(\omega(z) - \frac{\sigma \tau}{2}\right)}\nonumber\\
  &&\Theta_{22}(z)=(z-\lambda_1)^{-\frac{1}{2}} e^{\Delta(z)}\nonumber\\ &&\quad \quad
  \quad \times \frac{\theta_3\left(\omega(z)-\beta(\lambda)-\kappa +\frac{\sigma
  \tau}{2}\right)}{\theta_3\left(\omega(z) + \frac{\sigma
  \tau}{2}\right)},\label{thetad}
  \end{eqnarray}
  where  $\sigma = 1$ in  Case $1$ and $\sigma =0$ in Case 2,
  and $\beta(\lambda)$, $\omega(z)$ and $\kappa$ are  defined in
  Eqs.~(\ref{notion1}-\ref{notion2}).   The branch of
  $(z-\lambda_1)^{-\frac{1}{2}}$ is defined on the
  $z$-plane cut along the part of the line $\Sigma$ which
  is to the right of $\lambda_{1}\equiv \lambda_{A}$, and it is fixed by the
  condition
  $\arg (z-\lambda_1) = \pi, \quad \mbox{if}\quad  z - \lambda_{1} < 0$.

  The matrix function $\Theta(z)$ is defined on ${\Bbb C} \backslash
  \, {\Sigma}$. However, taking into account equations
(\ref{jinq0} - \ref{jinq2}) describing the jumps of
  the integrals $\omega(z)$ and $\Delta(z)$ over
  the line $\Sigma$ and  the
  properties (\ref{theta1}) and (\ref{theta2})
  of the theta function, one can see that $\Theta(z)$
  is actually extended to the analytic
  function defined on ${\Bbb C} \backslash \, \{ J_1\cup J_2\}$
which satisfies the jump relations
\begin{eqnarray}
  \Theta_+(z)=\Theta_-(z) \sigma_1\qquad z\in J_1 \label{J1jump}\\
  \Theta_+(z)=\Theta_-(z)\Lambda \sigma_1\Lambda^{-1}\quad
  z\in J_2.\label{J2jump}\\
  \Lambda=i\left(\begin{array}{cc}
                \lambda+1&0\\
                0&\lambda-1
                \end{array}\right),\quad \sigma_1=\left(\begin{array}{cc}
                0&1\\
                1&0
                \end{array}\right).
  \end{eqnarray}
The end points of the intervals $J_{1,2}$ are singular points
of the function $\Theta(z)$. Specifically, one can notice that
$$
\Theta(z) = O\left(\frac{1}{\sqrt{z - \lambda_{1}}}\right),
\quad \mbox{as}\quad z \sim \lambda_{1},
$$
and
$$
\Theta(z) = O\left(\frac{1}{\sqrt{z - \lambda^{-1}_{2}}}\right),
\quad \mbox{as}\quad z \sim \lambda^{-1}_{2}.
$$
The last estimate follows from the fact that, according
to (\ref{omegaA}) and (\ref{thetazeros}), the only zero of the
denominators of the right hand sides in (\ref{thetad}) is
the point $z = \lambda^{-1}_{2}$ (which is $\lambda_{B}$
in the Case 1, i.e. when $\sigma = 1$, and $\lambda_{C}$ in the Case 2,
i.e. when $\sigma = 0$). At the end points $z = \lambda^{-1}_{1}$
and $z = \lambda_{2}$ the function
$\Theta(z)$ is holomrphic with respect to the variables $\sqrt{z - \lambda^{-1}_{1}}$
and $\sqrt{z - \lambda_{2}}$, respectively, although
$\det \Theta(z)$ vanishes at these points. In more detail, the  following
representations take place at the points $z = \lambda_{A}$,
$\lambda_{B}$, $\lambda_{C}$, and $\lambda_{D}$ :
\begin{itemize}
\item in a neighborhood of $z =\lambda_{A}$, cut along $J_{1}$,
\begin{equation}\label{A}
\Theta(z) = \Theta_{A}(z)(z-\lambda_{A})^
{\left(\begin{array}{cc}
                -1/2&0\\
                0&0
                \end{array}\right)}\left(\begin{array}{cc}
                1&-1\\
                1&1
                \end{array}\right)\,\,;
\end{equation}
\item in a neighborhood of $z =\lambda_{B}$, cut along $J_{1}$,
\begin{equation}\label{B}
\Theta(z) = \Theta_{B}(z)(z-\lambda_{B})^
{\left(\begin{array}{cc}
                1/2 - \sigma&0\\
                0&0
                \end{array}\right)}\left(\begin{array}{cc}
                1&-1\\
                1&1
                \end{array}\right)\,\,;
\end{equation}
\item in a neighborhood of $z =\lambda_{C}$, cut along $J_{2}$,
\begin{equation}\label{C}
\Theta(z) = \Theta_{C}(z)(z-\lambda_{C})^
{\left(\begin{array}{cc}
                \sigma - 1/2&0\\
                0&0
                \end{array}\right)}\Lambda\left(\begin{array}{cc}
                1&-1\\
                1&1
                \end{array}\right)\Lambda^{-1}\,\,;
\end{equation}
\item in a neighborhood of $z =\lambda_{D}$, cut along $J_{2}$,
\begin{equation}\label{D}
\Theta(z) = \Theta_{D}(z)(z-\lambda_{D})^
{\left(\begin{array}{cc}
                1/2&0\\
                0&0
                \end{array}\right)}\Lambda\left(\begin{array}{cc}
                1&-1\\
                1&1
                \end{array}\right)\Lambda^{-1}.
\end{equation}
\end{itemize}
In these formulae, the left matrix multipliers $\Theta_{A}(z)$,
$\Theta_{B}(z)$, $\Theta_{C}(z)$, and $\Theta_{D}(z)$ are matrix -
valued functions holomorphic (with respect to the
variable $z$ itself) and invertible as matrices at the
points $z = \lambda_{A}$,
$z = \lambda_{B}$,  $z = \lambda_{C}$, and  $z = \lambda_{D}$,
respectively.
To prove, say, representation (\ref{A}) let us introduce the model function,
$$
\Theta_{0}(z) := (z-\lambda_{A})^
{\left(\begin{array}{cc}
                -1/2&0\\
                0&0
                \end{array}\right)}\left(\begin{array}{cc}
                1&-1\\
                1&1
                \end{array}\right).
$$
This function is well defined in a small disk centered at $\lambda_{A}$
and cut along $J_{1}$. By a simple direct computation we observe
that the function $\Theta_{0}(z)$ satisfies accross the portion
of $J_{1}$ lying in the disk {\it exactly} the same jump condition
as the one indicated in (\ref{J1jump}). Therefore, the matrix ratio,
$$
\Theta_{A}(z) := \Theta(z)\Theta^{-1}_{0}(z),
$$
does not have any jumps at all in the neighborhood of $\lambda_{A}$.
Hence, $z=\lambda_{A}$ is its isolated singular point. From the explicit formulae
(\ref{thetad}) it follows that, at worst,
$$
|\Theta_{A}(z)| \leq \frac{C}{|z-\lambda_{A}|^{1/2}},
$$
which implies that $z=\lambda_{A}$ is in fact a removable singularity.
Equation (\ref{A}) follows . The other equations from the list (\ref{A} - \ref{D})
can be proven in the similar way.

Strictly speaking, we have not completed the proof
of  equations (\ref{A} - \ref{D}). We need to show that
all the left multiplies are invertible matrices. To see this
we note that
  \begin{eqnarray}
  \Theta_{11}(\infty)&=&e^{\Delta_0}\frac{\theta_3\left(\beta(\lambda)+
  \frac{\sigma \tau}{2}\right)}
  {\theta_3\left(\kappa + \frac{\sigma \tau}{2}\right)}\label{tinf1}\\
  \Theta_{22}(\infty)&=&e^{\Delta_0}\frac{\theta_3\left(\beta(\lambda) -
  \frac{\sigma \tau}{2}\right)}
  {\theta_3\left(\kappa + \frac{\sigma \tau}{2}\right)}\label{tinf2}\\
  \Theta_{12}(\infty)&=&\Theta_{21}(\infty)=0\,\,,\label{tinf3}
  \end{eqnarray}
  and
  \begin{equation}\label{detTheta}
  \det \Theta(z) \equiv \phi(z) \det
  \Theta(\infty)\sqrt{\frac{\lambda_{2}}{\lambda_{1}}}.
  \end{equation}
To prove the latter equation, consider the ratio
$$
\psi(z) := \det \Theta(z)/\phi(z).
$$
The jump conditions (\ref{J1jump}) and (\ref{J2jump}) imply
that the function $\psi(z)$ has no jumps while the
representations (\ref{A} - \ref{D}) show that neither it
has poles. Hence, this function is a constant and we have that
$$
\psi(z) \equiv \psi(\infty) = \det
  \Theta(\infty)\sqrt{\frac{\lambda_{2}}{\lambda_{1}}}.
$$
This proves (\ref{detTheta}). In its turn, equation (\ref{detTheta})
implies the invertability of $\Theta_{k}(\lambda_{k})$ for
all $k = A, B, C, D$. In addition we also have that
$$
\det\Theta(z) \neq 0, \quad \mbox{for all}\quad z \notin
\{\lambda^{-1}_{1}, \lambda_{2}\}.
$$

 In addition to matrix-valued function $\Theta(z)$,
we introduce the matrix-valued function
  \begin{eqnarray}
  Q(z)=\left( \begin{array}{cc}
  \phi(z) & -\phi(z)\\
  i&i
  \end{array}     \right)\label{td}
  \end{eqnarray}
  Note that $Q(z)$ diagonalizes original jump matrix $\Phi(z)$:
  \begin{equation}\label{factoriz}
  \Phi(z)=Q(z)\Lambda Q^{-1}(z)
  \end{equation}
  and $Q(z)$ is analytic on ${\Bbb C}\backslash\, \{J_1\cup J_2\}$
  and
  \begin{equation}\label{Qjump}
  Q_+(z)=Q_-(z)\sigma_1, \quad z\in J_1\cup J_2.
  \end{equation}
Moreover, at the end points of the intervals $J_{1,2}$, which
are the only singularities of the matrix function $Q(z)$,
it behaves in the way similar to the behavior
of the function $\Theta(z)$, i.e. we have (cf. (\ref{A} - \ref{D})),
\begin{itemize}
\item in a neighborhood of $z =\lambda_{A}$, cut along $J_{1}$,
\begin{equation}\label{A0}
Q(z) = Q_{A}(z)(z-\lambda_{A})^
{\left(\begin{array}{cc}
                -1/2&0\\
                0&0
                \end{array}\right)}
{\left(\begin{array}{cc}
                1&-1\\
                1& 1
                \end{array}\right)}\,\,;
\end{equation}
\item in a neighborhood of $z =\lambda_{B}$, cut along $J_{1}$,
\begin{equation}\label{B0}
Q(z) = Q_{B}(z)(z-\lambda_{B})^
{\left(\begin{array}{cc}
                1/2 - \sigma&0\\
                0&0
                \end{array}\right)}{\left(\begin{array}{cc}
                1&-1\\
                1& 1
                \end{array}\right)}\,\,;
\end{equation}
\item in a neighborhood of $z =\lambda_{C}$, cut along $J_{2}$,
\begin{equation}\label{C0}
Q(z) = Q_{C}(z)(z-\lambda_{C})^
{\left(\begin{array}{cc}
                \sigma - 1/2&0\\
                0&0
                \end{array}\right)}{\left(\begin{array}{cc}
                1&-1\\
                1& 1
                \end{array}\right)}\,\,;
\end{equation}
\item in a neighborhood of $z =\lambda_{D}$, cut along $J_{2}$,
\begin{equation}\label{D0}
Q(z) = Q_{D}(z)(z-\lambda_{D})^
{\left(\begin{array}{cc}
                1/2&0\\
                0&0
                \end{array}\right)}{\left(\begin{array}{cc}
                1&-1\\
                1& 1
                \end{array}\right)},
\end{equation}
\end{itemize}
with some holomorphic and invertible at the respective points
matrices $Q_{A}(z)$, $Q_{B}(z)$, $Q_{C}(z)$, and $Q_{D}(z)$.
We are now ready to present  the solution $U_{\pm}(z)$
  of the Riemann-Hilbert problem (i-iii).
  Put
\begin{equation}A=Q(\infty)
  \Lambda^{-1}\Theta^{-1}(\infty),
\end{equation}
and define
  \begin{eqnarray} U_-(z):=A \Theta(z) \Lambda Q^{-1}(z),\quad |z|\ge 1 \label{u-d1}\\
  U_+(z):=Q(z)\Theta^{-1}(z) A^{-1},\quad |z|\le 1.\label{u-d}
  \end{eqnarray}
  By virtue of Eq.~(\ref{factoriz}), we only need
  to be sure that $U_-(z)$ and $U_+(z)$ are
  analytic and matrix invertible for $|z|>1$ and $|z|<1$ respectively.
  From the jump properties (\ref{J1jump} - \ref{J2jump}) of $\Theta(z)$ and
  (\ref{Qjump}) of $Q(z)$ it follows that
  $U_{\pm}$ have no jumps across $J_{1,2}$, and hence they might
  have only possible isolated singularities at $\lambda_{1,2},\lambda_{1,2}^{-1}$.
  The analyticity at these points follows immediately
  from the representations (\ref{A}  - \ref{D}) and
  (\ref{A0}  - \ref{D0}) which show that the
  singularities, which the functions $\Theta(z)$ and
  $Q(z)$ do have at the end points of the segments
  $J_{1,2}$, are canceled out in the products (\ref{u-d1})-(\ref{u-d}).

  The excluded values of $\lambda$ for which the above construction fails
  are $\lambda = \pm 1$ and, in view of Eq.~(\ref{detTheta}), the zeros of
  $\theta_3\left(\beta(\lambda)+\frac{\sigma \tau}{2}\right)$, i.e. (see
  (\ref{thetazeros})),
  \begin{equation}\label{zeros}
  \pm \lambda_{m}, \quad \lambda_{m} =
  \tanh \left(m + \frac{1-\sigma}{2}\right)\pi \tau_{0}, \quad m \geq 0,
  \end{equation}
  where,
  $$
  \tau_{0} = -i\tau = -i
  \frac{\int_{\lambda_{B}}^{\lambda_{C}}\frac{dz}{w(z)}}
  {\int_{\lambda_{A}}^{\lambda_{B}}\frac{dz}{w(z)}}>0 .
  $$

  We conclude this section by noticing that, using the standard reduction
  (see e.g. \cite{ww}) of an arbitrary elliptic integral to the
  canonical elliptic integrals, we can rewrite the above expression
  for $\tau_{0}$  as
\begin{equation}\label{tau0ellip}
\tau_0= I(k')/I(k).
\end{equation}
Here  $I(k)$ denotes the complete elliptic integral of the first kind,
$$
I(k) = \int_{0}^{1}\frac{dx}{\sqrt{(1-x^2)(1 - k^{2}x^{2})}},
$$
$k'=\sqrt{1-k^2}$, and
  \begin{eqnarray}
   k= \left \{ \begin {array} {c} \sqrt{(h/2)^2+\gamma^2-1}\; /\; \gamma ,
  \;\;\;\mbox{Case 1a} \\ [0.3cm]
  \sqrt{1 -\gamma^2 - (h/2)^2}\; / \;\sqrt{1-(h/2)^2},
\;\;\; \mbox{Case 1b}\\ [0.3cm]
         \gamma\; / \;\sqrt{(h/2)^2+\gamma^2-1} ,\;\;\; \mbox{Case 2}
 \end{array}
  \right.
    \label{mod}
  \end{eqnarray}

\section{The Proof of Theorem 1.  Evaluation of the Entropy.}

Denote
\begin{equation}\label{S0}
s(\lambda):=
\frac{1}{2\pi} \int_{|z| =
1}\mbox{trace}\, \left[\left(U_{+}'(z)U_{+}^{-1}(z)
+V_{+}^{-1}(z)V_{+}'(z)\right)\Phi^{-1}(z)\right]dz
\end{equation}
The proof of the theorem 1 will be achieved by showing that
\begin{equation}\label{finalMay}
s(\lambda) =
\frac{d}{d\lambda}
  \ln \left[ \theta_3\left(
  \beta(\lambda)+\frac{\sigma \tau}{2}\right)
  \theta_3\left(\beta(\lambda)-\frac{\sigma \tau}{2}\right)
  \right].
\end{equation}
From equation (\ref{uv2}) we obtain
\begin{equation}\label{S0eval1}
s(\lambda)=\frac{1}{2\pi} \int_{|z| = 1}\mbox{trace}\,
\left[U_{+}'(z)U_{+}^{-1}(z)\left(\Phi^{-1}(z)-
\sigma_3\Phi^{-1}(z)\sigma_3\right)\right]dz.
\end{equation}
Denote
$$\Psi(z):=\Phi^{-1}(z)-\sigma_3\Phi^{-1}(z)\sigma_3=
\frac{2}{1-\lambda^2}\left( \begin{array}{cc}
0 & -\phi(z)\\
\phi^{-1}(z)&0
\end{array}     \right).$$
From equation (\ref{u-d}), we have
$$ U_{+}^{-1}(z)=  A\Theta(z)Q^{-1}(z),\quad U'_{+}(z)=
Q'(z) \Theta^{-1}(z) A^{-1}+Q(z) (\Theta^{-1})'(z) A^{-1}.$$
and from equation (\ref{td}), we have $$ Q^{-1}(z)=\frac{1}{2}\left(
\begin{array}{cc}
\phi^{-1}(z) & -i\\
-\phi^{-1}(z)&-i
\end{array}     \right). $$
 Then
\begin{eqnarray}
s(\lambda)&=&\frac{1}{2\pi} \int_{|z| = 1}\mbox{trace}\,
\left[U_{+}'(z)U_{+}^{-1}(z)\Psi(z)\right]dz\nonumber\\
&=&\frac{i}{\pi(1-\lambda^2)} \int_{|z| = 1}\mbox{trace}\, \left[
  \Theta^{-1}(z)\frac{d}{dz} \Theta(z)\sigma_3
 \right]dz.\label{S0eval2}
\end{eqnarray}

Let
\begin{equation}\label{1}
\alpha(z): =
\mbox{trace}\, \left[
\Theta^{-1}(z) \frac{d}{dz}\Theta(z) \sigma_3
\right].
\end{equation}
Then, one can easily see   that the
jumps of $\alpha(z)$ across $J_{1}\cup J_{2}$ satisfy the
equation,
\begin{equation}\label{sigmajump}
\alpha_{+}(z) = -\alpha_{-}(z).
\end{equation}
From the explicit formulae (\ref{thetad}) for the $\Theta$-matrix we derive the
following asymptotic representation (actually, the Laurent series)
at $z=\infty$
\begin{equation}\label{thetaLaurent}
\Theta(z) = \Theta(\infty)\left(I + \frac{1}{z}{\Bbb M} + ...\right),
\end{equation}
where $\Theta(\infty)$ is the diagonal matrix defined in (\ref{tinf1})-(\ref{tinf1})
and the matrix coefficient ${\Bbb M}$ is given by the equations,
\begin{equation}\label{M11}
{\Bbb M}_{11} = \Delta_{1} + \frac{\lambda_{1}}{2}
+\frac{1}{c}\frac{\theta'_{3}\left(\kappa + \frac{\sigma\tau}{2}\right)}
{\theta_{3}\left(\kappa + \frac{\sigma\tau}{2}\right)}
- \frac{1}{c}\frac{\theta'_{3}\left(\beta(\lambda)+\frac{\sigma\tau}{2}\right)}
{\theta_{3}\left(\beta(\lambda)+\frac{\sigma\tau}{2}\right)},
\end{equation}
\begin{equation}\label{M22}
{\Bbb M}_{22} = \Delta_{1} + \frac{\lambda_{1}}{2}
+\frac{1}{c}\frac{\theta'_{3}\left(\kappa + \frac{\sigma\tau}{2}\right)}
{\theta_{3}\left(\kappa + \frac{\sigma\tau}{2}\right)}
+ \frac{1}{c}\frac{\theta'_{3}\left(\beta(\lambda)-\frac{\sigma\tau}{2}\right)}
{\theta_{3}\left(\beta(\lambda)-\frac{\sigma\tau}{2}\right)},
\end{equation}
\begin{equation}\label{M12}
{\Bbb M}_{12} = - e^{-2\Delta_{0}}
\frac{\theta_{3}\left(\beta(\lambda)-2\kappa +\frac{\sigma\tau}{2}\right)}
{\theta_{3}\left(\beta(\lambda)+\frac{\sigma\tau}{2}\right)}
\frac{\theta_{3}\left(\kappa +\frac{\sigma\tau}{2}\right)}
{\theta_{3}\left(\kappa - \frac{\sigma\tau}{2}\right)}
\end{equation}
\begin{equation}\label{M21}
{\Bbb M}_{21} = - e^{-2\Delta_{0}}
\frac{\theta_{3}\left(\beta(\lambda)+2\kappa -\frac{\sigma\tau}{2}\right)}
{\theta_{3}\left(\beta(\lambda)-\frac{\sigma\tau}{2}\right)}
\frac{\theta_{3}\left(\kappa +\frac{\sigma\tau}{2}\right)}
{\theta_{3}\left(\kappa - \frac{\sigma\tau}{2}\right)}.
\end{equation}
In the above equations, $\Delta_{1}$ is the coefficient of $z^{-1}$
term of the expansion of the integral $\Delta(z)$ at $z =\infty$
and $\theta'_{3}$ means the derivative of the theta function with respect
to its own argument (not the derivative with respect to $\lambda$ !),
i.e.
$$
\theta'_{3}(s) = \frac{d}{ds}\theta_{3}(s).
$$
In its turn, representation (\ref{thetaLaurent}) immediately
implies the following asymptotic behavior of $\alpha(z)$ as
$z \to \infty$,
\begin{equation}\label{sigmainf}
\alpha(z) = \frac{m(\lambda)}{z^2} + O\left(\frac{1}{z^3}\right),
\quad z \to \infty.
\end{equation}
where
$$
m(\lambda) = -\mbox{trace}\, \left[{\Bbb M}\sigma_{3}\right]
$$
\begin{equation}\label{beta}
= \frac{1}{c}\left(\frac{\theta'_{3}\left(\beta(\lambda)+
\frac{\sigma\tau}{2}\right)}{\theta_{3}\left(\beta(\lambda)+
\frac{\sigma\tau}{2}\right)} +
\frac{\theta'_{3}\left(\beta(\lambda)-
\frac{\sigma\tau}{2}\right)}{\theta_{3}\left(\beta(\lambda)-
\frac{\sigma\tau}{2}\right)}
\right).
\end{equation}
The representations (\ref{A}) - (\ref{D}) of the function $\Theta(z)$
at the end points show that at the
end points, i.e. at the points $\lambda_{1}$,  $\lambda_{2}$,
$\lambda^{-1}_{1}$,  $\lambda^{-2}_{2}$,
the function $\alpha(z)$ has at most square root singularities. This, together
with (\ref{sigmajump}) and (\ref{sigmainf}) yield the final
explicit formula for $\alpha(z)$ (cf. the derivation of equation
(\ref{detTheta}) given above):
\begin{equation}\label{sigmaexpl}
\alpha(z) = \frac{m(\lambda)}{w(z)}.
\end{equation}

Equation (\ref{sigmaexpl}) means that
equation (\ref{S0eval2}) can be rewritten as
\begin{equation}\label{s1}
s(\lambda)
= -\frac{2i}{\pi}
\frac{m(\lambda)}{1-\lambda^2}\int_{\lambda_{A}}^{\lambda_{B}}
\frac{dz}{w(z)},
\end{equation}
where the integration is taken on the left side of the cut
$[\lambda_{A}, \lambda_{B}]$.
By virtue of (\ref{beta}) and formula (\ref{important})
for the constant $c$, the last equation can be
transformed into
$$
s(\lambda)= -\frac{i}{\pi}\frac{1}{1-\lambda^2}
$$
\begin{equation}\label{s11}
\times \left(\frac{\theta'_{3}\left(\beta(\lambda)+\frac{\sigma\tau}{2}\right)}
{\theta_{3}\left(\beta(\lambda)+\frac{\sigma\tau}{2}\right)}
+\frac{\theta'_{3}\left(\beta(\lambda)-\frac{\sigma\tau}{2}\right)}
{\theta_{3}\left(\beta(\lambda)-\frac{\sigma\tau}{2}\right)}
\right).
\end{equation}
Taking into account that
$$
\frac{d}{d\lambda}\beta(\lambda) = -\frac{i}{\pi}\frac{1}{1-\lambda^2},
$$
we  arrive to the formula (\ref{finalMay}). Theorem 1 is proven.
\vskip .5in
We are now going to show that the two expressions for the entropy
$S(\rho_{A})$, i.e. equations (\ref{3333May}) and (\ref{eaaMay})
coincide. We will proceed by defining $S(\rho_{A})$ by equation
(\ref{eaaMay}) and showing that formula (\ref{3333May}) holds.
To this end we rewrite (\ref{eaaMay}) as
\begin{equation}\label{eaaMay1}
S(\rho_{A}) = \lim_{\epsilon\to 0^{+}}S_{\epsilon}(\rho_{A}),
\end{equation}
$$
S_{\epsilon}(\rho_{A}) = \lim_{L\to \infty} \frac{1}{4\pi \mathrm{i}}
  \oint_{\Gamma'} \mathrm{d} \lambda\,  e(1+\epsilon, \lambda)
  \frac{\mathrm{d}}{\mathrm{d} \lambda} \ln
  \left(D_{L}(\lambda)(\lambda^{2} -1)^{-L}\right),
$$
and observe that, by virtue of theorem 1,
$$
 \frac{1}{4\pi \mathrm{i}}
  \oint_{\Gamma'} \mathrm{d} \lambda\,  e(1+\epsilon, \lambda)
  \frac{\mathrm{d}}{\mathrm{d} \lambda} \ln
 \left(D_{L}(\lambda)(\lambda^{2} -1)^{-L}\right)
  =\frac{1}{4\pi \mathrm{i}}
  \oint_{\Gamma'} \mathrm{d} \lambda\,  e(1+\epsilon, \lambda)s(\lambda) + O(\rho^{-L}),
  $$
where $s(\lambda)$ is given by (\ref{finalMay}). Therefore,
\begin{equation}\label{May111}
S_{\epsilon}(\rho_{A})
=\frac{1}{4\pi \mathrm{i}}
  \oint_{\Gamma'} \mathrm{d} \lambda\,  e(1+\epsilon, \lambda)s(\lambda).
\end{equation}
The function $s(\lambda)$ is analytic on the $\lambda$ - plane minus
the points $\pm 1$ and $\pm \lambda_{m}$. Moreover, it is an odd function
with the zero at $\lambda =\infty$. Therefore, its Laurent series at infinity
is of the form,
$$
s(\lambda) = \frac{c_{3}}{z^3} + \frac{c_{4}}{z^4} + ...
$$
This implies that  (\ref{May111}) can be transformed as follows ( see Fig. \ref{fig1}).
$$
S_{\epsilon}(\rho_{A})
=\frac{1}{4\pi \mathrm{i}}
  \oint_{\Gamma'} \mathrm{d} \lambda\,  e(1+\epsilon, \lambda)s(\lambda)
$$
$$
= \lim_{R \to \infty}\frac{1}{4\pi \mathrm{i}}
  \oint_{\Gamma} \mathrm{d} \lambda\,  e(1+\epsilon, \lambda)s(\lambda)
$$

$$
=\frac{1}{4\pi i}\int_{-\infty}^{-1-\epsilon}
\left[-\frac{1+\epsilon +\lambda}{2}\left(\ln \left|\frac{1+\epsilon +\lambda}{2}\right|
+i\pi \right) + \frac{1+\epsilon +\lambda}{2}\left(\ln \left|\frac{1+\epsilon +\lambda}{2}\right|
-i\pi \right)\right]s(\lambda)d\lambda
$$
$$
+\frac{1}{4\pi i}\int_{1+\epsilon}^{\infty}
\left[-\frac{1+\epsilon -\lambda}{2}\left(\ln \left|\frac{1+\epsilon -\lambda}{2}\right|
-i\pi \right) + \frac{1+\epsilon -\lambda}{2}\left(\ln \left|\frac{1+\epsilon -\lambda}{2}\right|
+i\pi \right)\right]s(\lambda)d\lambda
$$

$$
= -\frac{1}{4}\int_{-\infty}^{-1-\epsilon}(1+\epsilon+\lambda)s(\lambda)d\lambda
+ \frac{1}{4}\int_{1+\epsilon}^{\infty}(1+\epsilon-\lambda)s(\lambda)d\lambda
$$

\begin{equation}\label{May112}
= \frac{1}{2}\int_{1+\epsilon}^{\infty}(1+\epsilon-\lambda)s(\lambda)d\lambda,
\end{equation}
where in the last equation we have again used the oddness of  the function
$s(\lambda)$.
Representing the function $s(\lambda)$  as
\begin{equation}\label{May113}
s(\lambda) =
\frac{d}{d\lambda}
  \ln \frac{ \theta_3\left(
  \beta(\lambda)+\frac{\sigma \tau}{2}\right)
  \theta_3\left(\beta(\lambda)-\frac{\sigma \tau}{2}\right)}
  {\theta_{3}^{2}\left(\frac{\sigma \tau}{2}\right)},
  \end{equation}
we can perform in (\ref{May112}) integration by parts. Indeed,
the function
$$
t(\lambda) \equiv
\ln \frac{ \theta_3\left(
  \beta(\lambda)+\frac{\sigma \tau}{2}\right)
  \theta_3\left(\beta(\lambda)-\frac{\sigma \tau}{2}\right)}
  {\theta_{3}^{2}\left(\frac{\sigma \tau}{2}\right)}
$$
is smooth for all $\lambda \in [1+\epsilon, +\infty)$, and
$$
t(\lambda) = O\left(\frac{1}{\lambda^2}\right), \quad \lambda \to \infty.
$$
This yields
the following, alternative to (\ref{May111}), integral representation
to the entropy $S_{\epsilon}(\rho_{A})$.
\begin{equation}\label{May114}
S_{\epsilon}(\rho_{A})
=\frac{1}{2}\int_{1+\epsilon}^{\infty}
\ln \frac{ \theta_3\left(
  \beta(\lambda)+\frac{\sigma \tau}{2}\right)
  \theta_3\left(\beta(\lambda)-\frac{\sigma \tau}{2}\right)}
  {\theta_{3}^{2}\left(\frac{\sigma \tau}{2}\right)}\,d\lambda,
  \end{equation}

It is an easy corollary of the second periodicity property (\ref{theta2May}) of
the theta function that
\begin{equation}\label{theta4May}
\ln \theta_{3}(ia + b) = \frac{\pi}{\tau_{0}}a^{2}\left(1 + O\left(\frac{1}{a}\right)\right),
\quad a\to \pm \infty,
\end{equation}
$$
\mbox{dist}\,\left(ia + b; \left \{\frac{1}{2} +\frac{\tau}{2} + n + m\tau \right\}\right) \geq \delta_{0} > 0.
$$
Observe that
$$
\beta(\lambda)  \in i{\Bbb R} \,\, \left(\mbox{mod}\,{\Bbb Z} \right),\quad
\forall \lambda \in (-\infty, -1)\cup (1, \infty),
$$
and hence
$$
\mbox{dist}\,\left(\beta(\lambda)
\pm  \frac{\sigma \tau}{2}; \left \{\frac{1}{2} +\frac{\tau}{2} + n + m\tau \right\}\right) \geq \frac{1}{2} > 0.
$$
Also,
$$
i\beta(\lambda) \to \pm \infty,
$$
as $\lambda  \to \pm 1 \pm 0$. Therefore, (\ref{theta4May}) is applicable and we see
that the function $t(\lambda)$ satisfies the estimate,
\begin{equation}\label{estMay111}
t(\lambda) \sim \frac{1}{2\pi \tau_{0}}\ln^{2} \left| \frac{\lambda -1}{\lambda + 1}\right|,
\quad \lambda \to \pm 1\pm 0,
\end{equation}
and hence  is integrable
at $\lambda = \pm 1$. This means, we can take limit $\epsilon \to 0$ in
(\ref{May114}) at arrive to the final integral formula for the
entropy $S(\rho_{A})$ (cf. equation (\ref{33May}))
\begin{equation}\label{May115}
S(\rho_{A})
=\frac{1}{2}\int_{1}^{\infty}
\ln \frac{ \theta_3\left(
  \beta(\lambda)+\frac{\sigma \tau}{2}\right)
  \theta_3\left(\beta(\lambda)-\frac{\sigma \tau}{2}\right)}
  {\theta_{3}^{2}\left(\frac{\sigma \tau}{2}\right)}\, d\lambda.
  \end{equation}

\begin{figure}[ht]
  \begin{center}
  \includegraphics[width=3in,clip]{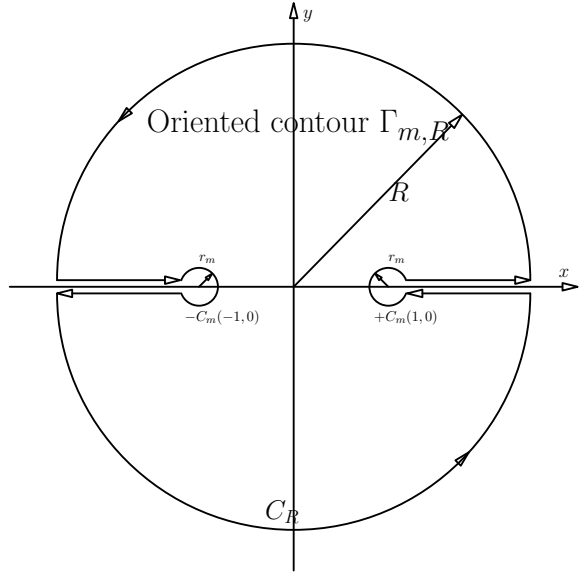}
  \end{center}
  \caption{\it Contours \mbox{$\Gamma_{m,R}$} with small circles $\pm C_{m}$ around
the points $\pm 1$ with the radius $ r_{m} = 1 - \left( \frac{1}{1 +
e^{\pi \tau_{0}}}\lambda_{m} +  \frac{e^{\pi \tau_{0}}}{1 + e^{\pi
\tau_{0}}}\lambda_{m+1}\right)$ and the big circle $C_{R}$ centered
at zero with radius $R$, the two copies of the interval $[1+r_{m},
R]$ - each lying on one of the sides of the cut from $1$ to
$+\infty$, and two similar copies of the symmetric interval, $[-R,
-1-r_{m}]$ } \label{fig3}
  \end{figure}

To obtain  the infinite sum representation (\ref{3333May}) we
introduce the positively oriented contour $\Gamma_{m, R}$ in
Fig~$\ref{fig3}$.

Then, by residue theorem,
$$
\sum_{k = -m -(1-\sigma)}^{m}e(1,\lambda_{k})
= \frac{1}{4\pi \mathrm{i}}
  \oint_{\Gamma_{m,R}} \mathrm{d} \lambda\,  e(1, \lambda)s(\lambda)
 $$
 $$
 = \lim_{R \to \infty} \frac{1}{4\pi \mathrm{i}}
  \oint_{\Gamma_{m,R}} \mathrm{d} \lambda\,  e(1, \lambda)s(\lambda).
$$
The last integral can be transformed (cf. derivation of (\ref{May112}))
to the expression
$$
 \frac{1}{4\pi \mathrm{i}}
  \oint_{C_{m}} \mathrm{d} \lambda\,  e(1, \lambda)s(\lambda)
  +  \frac{1}{4\pi \mathrm{i}}
  \oint_{-C_{m}} \mathrm{d} \lambda\,  e(1, \lambda)s(\lambda)
$$
$$
+\frac{1}{4\pi i}\int_{-\infty}^{-1-r_{m}}
\left[-\frac{1+\lambda}{2}\left(\ln \left|\frac{1 +\lambda}{2}\right|
+i\pi \right) + \frac{1 +\lambda}{2}\left(\ln \left|\frac{1 +\lambda}{2}\right|
-i\pi \right)\right]s(\lambda)d\lambda
$$
$$
+\frac{1}{4\pi i}\int_{1+r_{m}}^{\infty}
\left[-\frac{1 -\lambda}{2}\left(\ln \left|\frac{1 -\lambda}{2}\right|
-i\pi \right) + \frac{1 -\lambda}{2}\left(\ln \left|\frac{1 -\lambda}{2}\right|
+i\pi \right)\right]s(\lambda)d\lambda
$$

$$
=\frac{1}{4\pi \mathrm{i}}
  \oint_{C_{m}} \mathrm{d} \lambda\,  e(1, \lambda)s(\lambda)
  +  \frac{1}{4\pi \mathrm{i}}
  \oint_{-C_{m}} \mathrm{d} \lambda\,  e(1, \lambda)s(\lambda)
$$
$$
 -\frac{1}{4}\int_{-\infty}^{-1-r_{m}}(1+\lambda)s(\lambda)d\lambda
+ \frac{1}{4}\int_{1+r_{m}}^{\infty}(1-\lambda)s(\lambda)d\lambda
$$

\begin{equation}\label{May118}
= \frac{1}{4\pi \mathrm{i}}
  \oint_{C_{m}} \mathrm{d} \lambda\,  e(1, \lambda)s(\lambda)
  +  \frac{1}{4\pi \mathrm{i}}
  \oint_{-C_{m}} \mathrm{d} \lambda\,  e(1, \lambda)s(\lambda)
+\frac{1}{2}\int_{1+r_{m}}^{\infty}(1-\lambda)s(\lambda)d\lambda,
\end{equation}
By a straightforward calculation we find that uniformly
in $\lambda \in C_{m}$ the following estimates hold,
$$
\beta(\lambda) = -\frac{\vartheta}{2\pi}
-\tau \left(m+1 -\frac{\sigma}{2}\right) + O\left(e^{-2\pi \tau_{0}}\right),
\quad \lambda = 1 +r_{m}e^{i\vartheta}, \quad m \to \infty,
$$
and hence
$$
\mbox{dist}\,\left(\beta(\lambda)
\pm  \frac{\sigma \tau}{2}; \left \{\frac{1}{2} +
\frac{\tau}{2} + n + m'\tau \right\}\right) \geq \frac{\tau_{0}}{4} > 0,
$$
for all $\lambda \in \pm C_{m}$ and sufficiently large $m$.
This again implies the applicability of (\ref{theta4May}) and, as a consequence,
the estimate
$$
|s(\lambda)| \leq \frac{C}{r_{m}}\ln \frac{1}{r_{m}}, \quad \lambda \in \pm C_{m}.
$$
The last estimate in its turn means that the integrals over $\pm C_{m}$ in
(\ref{May118}) vanish as $m \to \infty$ while the  integral over $[1+r_{m}, \infty]$
becomes the integral over $[1, \infty]$. Therefore we arrive at the relation,
$$
\sum_{k = -\infty}^{\infty}e(1,\lambda_{k})
\equiv \sum_{m=-\infty}^{\infty}H(\lambda_{m})
\equiv \sum_{m = -\infty}^{\infty}(1 + \lambda_{m})\ln \frac{2}{1+\lambda_{m}}
$$
$$
=\frac{1}{2}\int_{1}^{\infty}(1-\lambda)s(\lambda)d\lambda
= \frac{1}{2}\int_{1}^{\infty}
\ln \frac{ \theta_3\left(
  \beta(\lambda)+\frac{\sigma \tau}{2}\right)
  \theta_3\left(\beta(\lambda)-\frac{\sigma \tau}{2}\right)}
  {\theta_{3}^{2}\left(\frac{\sigma \tau}{2}\right)}\, d\lambda,
$$
which completes our evaluation of the entropy.

\section{Some critical cases.}

  The entropy has singularities at {\it phase transitions}. When
  $\tau \to 0$ we can use   Landen
  transform (see  \cite{ww}) to get the following estimate
  of the theta-function for small
  $\tau$ and pure imaginary $s$:
  $$\ln \frac{\theta_{3}\left(s \pm
  \frac{\sigma \tau}{2}\right)}{\theta_{3}\left(\frac{\sigma \tau}{2}\right)} =
  \frac{\pi}{i\tau}s^{2} \mp \pi i
  \sigma s + O\left(\frac{e^{-i\pi/\tau}}{\tau^{2}} s^2\right), ~\textrm{as $\tau \to
  0$}.$$
  Now the leading term in the expression for  the entropy (\ref{May115})
   can be replaced   by
  \begin{equation}\label{4}
  S(\rho_{A}) = \frac{\mathrm{i}\pi}{6\tau}+
   O\left(\frac{e^{-i\pi/\tau}}{\tau^{2}}\right)
  \quad \textrm{for
  $\tau \to 0$}.
  \end{equation}
  Let us consider two physical situations corresponding to small
  $\tau$ depending on the case defined on the page 2:

  \begin{enumerate}
  \item{\it Critical magnetic field}:  $\gamma\neq 0$ and $h\to 2$.

  This  is included  in our   Case
   $1$a  and  Case $2$, when $h> 2\sqrt{1-\gamma^2 }  $.
  As $h\to 2$ the end points of the cuts  $\lambda_B \to \lambda_C$, so $\tau$
  given by Eq.~(\ref{important}) simplifies and
   we obtain from Eq.~(\ref{4}) that the entropy is very large:
  \begin{eqnarray} &S(\rho_{A}) = -\frac{1}{6} \ln |2-h| + \frac{1}{3}\ln
  4\gamma \Longrightarrow  +\infty \label{cardy} \\
  &~~h\longrightarrow 2 \qquad \mbox{and} \qquad  \gamma\neq 0 . \nonumber
  \end{eqnarray}
  Next correction is $O(|2-h|\ln^{2}|2-h|) $. This limit agrees with predictions of
  conformal approach \cite{cardy}.
   The
   first term in the right hand side of (\ref{cardy})
  can be represented as $(1/6)\ln \xi$, this confirms a conjecture of
   \cite{cardy}.  The correlation length $\xi$
    was  evaluated in \cite{barouch, mccoy}.

  \item{\it An approach to $XX$ model}:  $\gamma\to 0$ and $h<2$: It
  is included in Case $1$b, when  $0<h<2\sqrt{1-\gamma^2} $. Now
   $\lambda_B \to \lambda_C$ and  $\lambda_A \to \lambda_D$, we can calculate $\tau$
  explicitly. The entropy increases without a bound:
  \begin{eqnarray}
  &S^{0}(\rho_{A}) = -\frac{1}{3} \ln \gamma + \frac{1}{6}{\ln
  (4-h^2)}+\frac{1}{3}\ln 2 \Longrightarrow  +\infty , \nonumber\\
  &\gamma\longrightarrow  0 \qquad \mbox{at} \qquad  h<2
  \end{eqnarray}
  correction is $O(\gamma\ln^2 \gamma) $. This agrees with \cite{jin}.
  \end{enumerate}

  It is interesting to compare this critical behavior to Lipkin-Meshkov-Glick
  model. It is similar to $XY$ model but each pair of spins interact with equal
  force, one can say that it is a model on a complete graph. The critical behavior
  in  Lipkin-Meshkov-Glick was described in \cite{julien}, it is similar to $XY$,
  but actual critical exponents are different.    Lipkin-Meshkov-Glick  model  displays interesting behavior of the entropy. Both the concurrence and the entropy were studied  numerically and analiticaly
  [in the thermodynamical limit] in   \cite{jvid}.

  {\bf Remark }. {\it The zeros  $\lambda_{m}$  satisfy an estimate:
  $$
  |\lambda_{m+1} - \lambda_{m}| \leq 4\pi \tau_{0} \quad \mbox{with} \quad \tau_{0} =
  -i\tau.
  $$
  This means that  $(\lambda_{m+1} - \lambda_{m}) \to 0$ as $\tau \to 0$
  for every $m$. This is useful for understanding of  large $L$ limit of
  the $XX$ case
  corresponding  to $\gamma \to 0$, as  considered in \cite{jin}.  The estimate
  explains why in the $XX$ case  the singularities of the
  logarithmic derivative of the Toeplitz determinant
   $d\ln D_L(\lambda)/ d\lambda $ form
  a cut along the interval $[-1, 1]$,
  while in the $XY$ case it has  a discrete set of poles at points $\pm \lambda_{m}$
  of Eq.~(\ref{zeros}).}

  It would be interesting to generalize our approach to the a new class
  of quantum spin chains introduced recently by J.  Keating and F. Mezzadri,
  while study matrix models
   \cite{keat}.

   {\it Acknowledgments.} We would like to thank    P.Deift, B.McCoy,
   I.Peschel, F.Franchini and H.Widom
  for useful  discussions. This work was supported by NSF Grants DMS 0503712,
  DMR-0302758 and DMS-0401009. The first co-author
  thanks B. Conrey, F. Mezzardi, P. Sarnak, and N. Snaith - the organizers of the
  2004 program at the Isaac Newton Institute
  for Mathematical Sciences on Random Matrices, where part of this work
  was done, for an extremely stimulating research environment
  and hospitality during his visit.

\end{document}